**A DNA turbine powered by a transmembrane potential across a nanopore**

Xin Shi,[1, ‡] Anna-Katharina Pumm,[2] Christopher Maffeo,[3] Fabian Kohler,[2] Elija Feigl,[2] Wenxuan Zhao,[1] Daniel Verschueren,[1,†] Ramin Golestanian,[4, 5] Aleksei Aksimentiev,[3,*] Hendrik Dietz,[2,*] Cees Dekker[1,*]

1 Department of Bionanoscience, Kavli Institute of Nanoscience Delft, Delft University of Technology, Delft, The Netherlands.
2 Department of Bioscience, School of Natural Sciences, Technical University of Munich, Garching 85748, Germany & Munich Institute of Biomedical Engineering, Technical University of Munich, Boltzmannstraße 11, 85748 Garching, Germany.
3 Department of Physics, University of Illinois at Urbana-Champaign, Urbana, Illinois 61801, United States
4 Max Planck Institute for Dynamics and Self-Organization, 370777, Göttingen, Germany
5 Rudolf Peierls Centre for Theoretical Physics, University of Oxford OX1 3PU, Oxford, UK
‡ Current address: Department of Chemistry, KU Leuven, Leuven, Belgium.
† Current address: The SW7 Group, 86/87 Campden Street, W8 7EN, London, UK.
* Corresponding author email addresses: c.dekker@tudelft.nl, dietz@tum.de, aksiment@illinois.edu

**Abstract**

Rotary motors play key roles in energy transduction, from macroscale windmills to nanoscale turbines such as ATP synthase in cells. Despite our capabilities to construct engines at many scales, developing functional synthetic turbines at the nanoscale has remained challenging. Here, we experimentally demonstrate rationally designed nanoscale DNA-origami turbines with three chiral blades. These DNA nanoturbines are 24-27 nm in height and diameter and can utilise transmembrane electrochemical potentials across nanopores to drive DNA bundles into sustained unidirectional rotations of up to 10 revolutions/s. The rotation direction is set by the designed chirality of the turbine. All-atom molecular dynamics simulations show how hydrodynamic flows drive this turbine. At high salt concentrations, the rotation direction of turbines with the same chirality is reversed, which is explained by a change in the anisotropy of the electrophoretic mobility. Our artificial turbines operate autonomously in physiological conditions, converting energy from naturally abundant electrochemical potentials into mechanical work. The results open new possibilities for engineering active robotics at the nanoscale.

**Introduction**

At the heart of any active mechanical system is an engine which converts one type of energy, typically chemical or electrical, into mechanical work. In biological systems, such work is done by motor proteins such as kinesin[1], the bacterial flagella motor[2], and $F_oF_1$-ATP synthase[3,4]. In the latter, electrochemical potential energy from a concentration gradient of ions is converted into mechanical rotary motion of the $F_o$ motor, which drives the $F_1$ rotary complex to catalyse the synthesis of ATP, the molecule that provides free energy for many cellular processes. Despite the extensive knowledge and success of building rotary engines of sizes spanning many orders of magnitude on the macroscale, designing, building, and demonstrating functioning artificial nanoscale counterparts of these sophisticated biological motors has proven challenging.

The critical step of building such nanoscale rotary engines is to demonstrate their ability to transduce local free energy continuously and autonomously into designed mechanical motion and useful work. Previous work led to multiple designs of rotary assemblies[5-8], and established a certain level of directed motion as an external operator would manually cycle environmental conditions[9] such as light and temperature[10,11], chemical compounds[12,13], or alternate macroscale electric fields[14]. Molecular dynamics (MD) simulations have shown the conceptual feasibility of using a DNA helix to convert the electric field into torque[15]; however, experimental demonstration of a rotary mechanism programmed for sustained conversion of a transmembrane electric potential into the mechanical rotation has not been achieved, until now.

Here, we demonstrate a bottom-up designed DNA nanoturbine that is powered by a nanoscale hydrodynamic flow inside a nanopore. It contains a central axle decorated with three blades arranged in a chiral configuration, either left- or right-handed. The turbine has a height of 24 or 27 nm, comparable to the 20-nm tall ATP synthase. The turbine's stator is provided by a solid-state nanopore in a 20-nm thin silicon nitride membrane. Using single-molecule fluorescence, we monitor the rotation of the nanoturbine driven by either a DC voltage or a transmembrane ion gradient, which mimics the working environment of rotary motors in biological cells. As we demonstrate below, the nanoturbine can drive a long DNA bundle as a hydrodynamic load into sustained rotary motion up to 10 revolutions/s, equivalent to delivering tens of pN·nm torque, which compares well with the ~50 pN·nm torque that can be generated by natural ATP synthase[16,17].

**Design of DNA nanoturbines**

Our DNA nanoturbine is a multilayer DNA origami structure containing an intentionally designed chiral twist (Fig. 1a and 1b). The structure consists of 30 double-stranded (ds-)DNA helices, each 72 base pairs (bp) in length, on average, where the six parallel central helices form an axle, and the three 8-helix blades are obliquely attached to the axle and symmetrically spaced at 120 degrees angles across the circumference. The chiral twist in the blades of the turbine is induced by adjusting the number of base pairs between each staple crossover away from the 1-per-7-bp value required for an achiral structure[18], yielding a strongly right-handed twisted structure for an 8 bp crossover density and a left-handed twisted turbine for 6.5 bp spacings on average, while maintaining a good folding yield of the structures (see Supplementary Fig. S1-S5). The objects were self-assembled as described previously[19] (for details, see Methods). We used single-particle cryogenic electron microscopy (cryo-EM) to determine 3D electron density maps of the right-handed and the left-handed turbine structures (Fig. 1c, 1e and Fig. S6-S9). The cryo-EM reconstructions showed the desired structural features such as the three blades and the axel, and revealed the twisted orientation of the blades. The twist and the blade angles were measured from the cryo-EM data, yielding a -1.1 degree/bp twist density and a blade angle with respect to the turbine axis of -36 degree for the right-handed structure, and +0.69 degree/bp and +24 degree for the left-handed structure (Fig. 1d and 1f, details see Methods and Fig. S10). The turbine structures were 27 nm and 24 nm (right- and left-handed, respectively) tall and had diameters of 27 nm and 25 nm, respectively.

**Unidirectional rotation of DNA turbines driven by a salinity gradient or transmembrane voltage**

To demonstrate that the turbines can generate torque and work, we docked the structure into a nanopore, and optically monitored rotations at the single-molecule level. To create a hydrodynamic load as well as to hold the turbine in the nanopore and to facilitate optical tracking using super-resolution microscopy, we attached a 300 nm long DNA six-helix bundle as a crossbar featuring a reinforced 220-nm long central 16 helix bundle segment (SI Fig. S12) to the top part of the turbine axle as one continuous rigid body (Fig. 1g). One end of the DNA bundle was labelled with ten Cy3 fluorophores to allow continuous monitoring of its motion by fluorescence microscopy at 5-millisecond temporal resolution and sub-diffraction-limit localisation precision (see Methods). A 900-nm long loop of nicked dsDNA was engineered to extend from the bottom of the axle and act as a leash guiding the insertion of the turbine into the nanopore during the docking process (Fig. 1g). Coarse-grained simulations (SNUPI[20]) were used to analyse the structural rigidity of the integrated design (Fig. 1h). Proper folding of the entire assembly was verified by negative-stained transmission electron microscopy (TEM, Fig. 1i, for details see Methods). An array of 50-nm-diameter nanopores was fabricated in 20-nm thin silicon nitride membranes using electron-beam lithography and reactive ion etching (see Methods) and characterised with TEM (Fig. S13).

Our single-molecule observations show that the DNA turbine can drive a unidirectional rotation of the load under a transmembrane gradient of ion concentration. Initially, both compartments of the flow cell were filled with 50 mM NaCl buffer, and the DNA turbines were added to the *cis* compartment. Subsequently, a higher concentration of NaCl (0.5 to 3M) was flushed into the *trans* compartment (Fig. 2a), causing the DNA turbines to move towards the nanopores by diffusiophoresis, which led to the insertion of the turbines into the nanopores. As the leash guided the initial insertion of the large structure[21], the turbine oriented itself upon docking in the designed orientation, where the extended crossbar bundle prevented the turbine from translocating through the nanopore. After docking, we tracked the rotary motion of the DNA bundle by monitoring the position of the fluorophores that were located at one end of the bundle (Fig. S14).

As shown in Fig. 2, we observed clear, directed rotation of the turbine particles. Figure 2b shows the rotary motion of the bundle end, which can be traced to follow a circular path over time. Figure 2c shows the accumulative angular displacement corresponding to the example in Fig. 2b, which continues over 200 clockwise rotations over the full period of observation (40 s in this case). Fig. 2d and 2e show typical data for 228 turbines, where right- and left-handed turbines led to, respectively, upwards and downwards linear angular rotation curves. The linearity of these curves (and the corresponding superlinear mean-square angular rotation curves; Fig. S15) are direct evidence of a driven motion. The data show that the DNA turbines can exert a substantial torque onto the DNA load bundle and drive it into sustained unidirectional rotary motion.

Importantly, we find that the designed chirality sets the rotation direction. The turbine variant with left-handed turbine blades displayed preferentially counter-clockwise rotations (as viewed from the *cis* side). In contrast, the turbines with right-handed turbine blades showed almost exclusively clockwise rotations. These data indicate that the designed chirality controls the rotation direction. We determined the average angular velocity in different salt gradient conditions (see Fig. 2f and Fig. S16). The velocity directions corresponded well with the designed chirality of the structures, and the angular velocities were distributed with noticeable spread with maximum values as high as ~10 revolutions per second. We attribute the spread in the velocity distribution to heterogeneity in the local interactions between the DNA structure and the silicon nitride surface and to potential deformations in the DNA turbine crossbar that can modulate the rotation speed [22].

Subsequently, we operated the DNA turbines under a transmembrane voltage that was applied across the compartments at an equal 50 mM NaCl concentration. Immediately after applying the transmembrane voltage (100 mV, Fig. 3a and 3b), docking and rotary motion of DNA turbines was observed – see Fig. 3b and 3h for typical traces of a left-handed and a right-handed DNA turbine, respectively. Clear circular trajectories were obtained that indicated a sustained and constant rotary motion over time, very similar to the trajectories observed in the ion-gradient-driven experiments. The extracted rotational velocities of the DNA load bundle showed driven rotary motion with, again, predominantly the same rotation direction depending on the chirality of the turbine variant under study, as can be seen in Fig. 3c and 3i. From the rotational speed of the DNA beam, we estimated the torque of our DNA turbine (Supplementary Note 1 and Fig. S20) of tens of pN·nm. As a control, we also tested an approximately non-chiral, straight version of the turbine with the same crossbar load (Fig. S18), which was assembled by removing the residual twist in the blades [18]. For this variant, no preferred rotational directionality was observed, while some residual rotation without preferred directionality was observed due to the self-organisation of the DNA crossbar[21] (Fig. S18).

**Direction reversal under a transmembrane voltage**

To our surprise, we found that we can control the rotational direction of the DNA turbines by the ionic strength of the buffer. When the voltage-driven experiments were performed in a high-salt buffer containing 3 M NaCl instead of dilute 50 mM, the directionality of the rotary motion reversed for the same turbine. This phenomenon occurred for both turbine chiralities, as shown in Fig. 3d, e and 3j, k (and corresponding MSD plots in Fig. S19). By titrating the salt concentration from 50 mM to 3 M, we observed that the average rotation speed changed from positive to negative values for the left-handed turbines, i.e., rotations changed from counter-clockwise to clockwise directions (and a reverse crossover occurred for right-handed turbines), with a crossover at ~0.5-1M for both chiralities – see Fig. 3g, m (and corresponding histograms for the speed distribution in Fig. S21-S22). These observations point to the appearance of strong ionic effects on the water flow when the turbine operates in a high-ionic strength environment.

To understand this rotational reversal induced by different ionic strengths, we first sought insights from continuum theory. As a model for the turbine blade, we consider a rigid cylindrical DNA rod that is held at a fixed vertical position in a wide nanopore, orientated at an angle of θ with respect to the Z axis (cf. Schematic S1 in SI). This rod has a hydrodynamic mobility $M_h$ that is anisotropic as a rod moves twice faster through a liquid along its length than perpendicular to it[23]. Similarly, the electrophoretic mobility $M_{el}$ of the rod, which describes its freely suspended motion in an applied electric field $\boldsymbol{E}$, is anisotropic[23]. Notably, $\boldsymbol{M_{el}}$ is controlled by the electrical double layer, which changes with the ionic strength of the solution [24,25]. The combination of $\boldsymbol{M_h}$ and $\boldsymbol{M_{el}}$ determines the inplane force on a turbine blade in a nanopore, which will drive the rotation. We can derive an expression for the in-plane velocity component $v_x$ (see Supplementary Text 2) as follows,

$$v_x = \frac{\frac{1}{2} M_{el,\parallel} E_z \sin 2\theta}{\cos^2\theta + \left(\frac{M_{h,\perp}}{M_{h,\parallel}}\right)\sin^2\theta} \left(\frac{M_{h,\perp}}{M_{h,\parallel}} - \frac{M_{el,\perp}}{M_{el,\parallel}}\right).$$

This equation indicates that the sign of $v_x$, i.e. the rotation direction, is determined by the difference between the hydrodynamic and electrophoretic anisotropy ratios. While the hydrodynamic anisotropy ratio $M_{h,\perp}/M_{h,\parallel}$ is constant at a value of 0.5, the electrophoretic anisotropy ratio $M_{el,\perp}/M_{el,\parallel}$ can adopt values between 0 and 1, depending on ion concentration and DNA surface charge[25], e.g., $M_{el,\perp}/M_{el,\parallel}$ is decreasing from 1 to 0.5 with decreasing ion concentration for moderate surface charges, and adopting even smaller values for high surface charges[25]. Continuum theory thus indicates that a sign reversal of the rotations may occur due to a change in the electrophoretic anisotropy ratio with salt concentration.

To elucidate the microscopic mechanism of the torque generation, we performed all-atom molecular dynamics (MD) simulations of the DNA origami turbine submerged in a low salt electrolyte solution, Fig. 4a. The application of a 100 mV/nm axial electric field was observed to rotate the turbine by ~120° in 56 ns in the expected direction (left-handed about the applied field axis, Fig. 4b and Fig. S24, see Methods for details). A water flow induced by a pressure gradient was observed to rotate the turbine in the expected direction, whereas reversing the direction of the electric field or the pressure gradient reversed the rotation direction. The turbine's rotation speed was found to be determined by the velocity of water molecules moving past the blades of the turbine, see Fig. S25b – all indicating a rotation mechanism similar to that of a macroscopic turbine. However, in equivalent simulations of the same DNA turbine carried out in 3M NaCl, the direction of rotation reversed, while the overall rotation speed decreased, see Fig. 4d, e. Furthermore, the average effective torque produced by the turbine under high salt was seen to change sign compared to the torque under a low salt (Fig. 4f). These MD simulation results thus present a striking qualitative resemblance of the experimental results.

To understand the ion-concentration-dependent reversal of the effective torque, we simulated a DNA duplex that was orientated parallel or perpendicular (Fig. S26a) to an applied electric field. The observed electrophoretic anisotropy ratio $M_{el,\perp}/M_{el,\parallel}$ indeed changed with the measured salt concentration, from about 1 at 2.6M NaCl to 0.38 at 4 mM NaCl (Fig. 4h and Fig. S26b) – a very significant change that would reverse the direction of rotation according to the continuum model.

To determine the microscopic mechanism of the rotation reversal, we simulated a DNA duplex that was tilted 35° relative to the vertically applied field (Fig. 4g), which approximates the inclination of the blades in our turbines. While the duplex' centre of mass and the spinning angle were harmonically restrained, we measured the effective in-plane force $F_x$ acting on the duplex, Fig. 4i. At low salt, a negative force was measured – as expected – whereas at high salt, the average in-plane force was reversed to a positive value. As the flow profile around the DNA (Fig. 4j) is driven by the distribution of charges in the solvent, we calculated a 3D map of the solvent forces— the position-dependent average force on solvent voxels applied by the electric field (see Methods). The map revealed a slight overcharging of the DNA in the 3M condition (see Fig. 4k), which was nevertheless sufficient to substantially alter the flow of solvent near the DNA (see Fig. 4j). To prove the causal role of the ion distribution, we applied the high/low salt solvent force on the low/high salt systems which, gratifyingly, swapped the effective force distributions, see Fig. 4l.

**Conclusions**

In summary, we have demonstrated a new type of autonomous active nanomachine, a DNA nanoturbine. We showed its functionality from the observation of a sustained rotation of a DNA load bundle when the turbines were driven by an electrophoretic or hydrodynamic flow of solvent through a nanoscale pore. Building from our previous work [22,26], the direction of rotation is now controlled by the designed chirality of the turbine blades as well as by the ionic strength of the buffer. The latter revealed, strikingly, that ionic interactions can even reverse the rotation direction, as the electrophoretic mobility anisotropy changes with buffer salt concentration. Up to tens of pN·nm torque can be generated by these merely ~25-nm tall turbines (an order of magnitude smaller than previously reported DNA origami rotors) [22], numbers that are comparable to the $F_o$ motor of ATP synthase [17]. The DNA turbines could operate autonomously on physiologically and biologically compatible energy sources (electrochemical potentials) and work at a scale comparable to nature's own motor proteins, without requiring manual cyclic intervention.

Our work demonstrates a practical approach to designing nanoscale active engines, i.e., by using chiral nanoscale structures to leverage transmembrane potential differences through nanoscale hydrodynamic interactions[15]. We believe this to be a powerful new approach to building active nanoscale systems, in particular, because of the extensible design with DNA nanotechnology, its flexibility and potential for integration with various bio-compatible membrane systems, and its potential compatibility to physiological environments. As in ATP synthase, microscopic reversibility may be exploited in future variants of such turbines to couple this mechanical rotation to uphill chemical synthesis. Nanoscale hydrodynamic turbines constitute biocompatible engines that may serve as a first step towards building self-powered nanorobotic systems in molecular biology-relevant environments. Current limitations of our apporach include the non-specific interaction between the turbine and the solid surface, the wide variance of the rotational speed, and the finite power efficiency. Further work can be done to demonstrate nanorobotic systems based on the turbine concept, specifically the integration of nanoscale motors into biocompatible membranes, improving their efficiency in utilising ion gradients and integrating them with other passive functional nanomachines to perform more complex tasks.

**Acknowledgements**

We thank Miloš Tišma for help with fluorescence microscopy imaging, Philipp Ketterer for initial turbine designs. We acknowledge funding support by the ERC Advanced Grant no. 883684 and the NanoFront and BaSyC programs (to C.D). This work was further supported by an ERC Consolidator Grant to H.D. (GA no. 724261), the Deutsche Forschungsgemeinschaft via the Gottfried-Wilhelm-Leibniz Program (to H.D.) and the SFB863 Project ID 111166240 TPA9 (to H.D.). A.A. and C.M. acknowledge support through National Science Foundation (USA) under grant DMR-1827346 (to A.A.). This work has received support from the Max Planck School Matter to Life and the MaxSynBio Consortium, which are jointly funded by the Federal Ministry of Education and Research (BMBF) of Germany, and the Max Planck Society (to R.G.). Supercomputer time was provided through Leadership Resource Allocation MCB20012 on Frontera and through ACCESS allocation MCA05S028.

**Author contributions**

HD and CD conceived the concept of DNA turbines in nanopores and nanopore arrays. AKP and XS co-designed the geometries of the turbines. AKP designed and prepared the DNA origami structures. XS designed the nanopore experiment and fabricated nanopore devices, supported by DV. XS and WZ conducted nanopore experiments. FK performed Cryo-EM measurements, EF performed EM data analysis and the atomic model construction. XS and DV wrote the data analysis scripts and analysed data. RG constructed the continuum model. CM and AA designed and conducted MD simulations. All authors discussed the findings and co-wrote the manuscript.

**Competing interests**

The authors declare that they have no competing interests.

# Figures

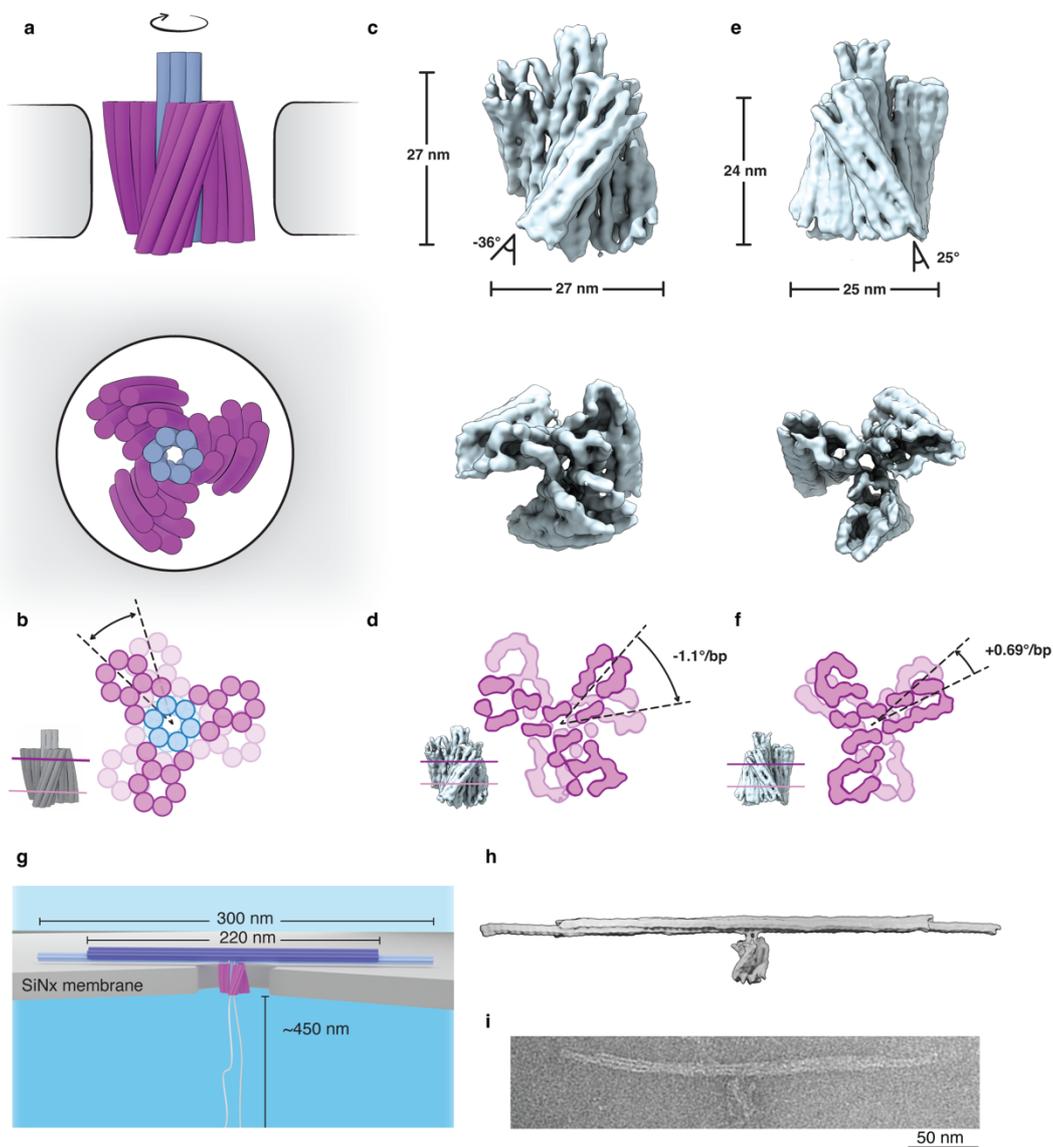

**Figure 1. Design of a nanopore-powered DNA origami turbine.**
**a.** Schematic of a right-handed DNA turbine docked into a nanopore (side view on top; axial view at bottom). **b.** Two cross sections of the DNA turbine highlighting the designed twist. **c.** 3D electron density map of the right-handed DNA turbine determined via single particle cryo-EM (Side and bottom view; see also Fig. S2). **d.** Cross sections at the top and the bottom of the 3D cryo-EM reconstruction right-handed turbine, highlighting the right-handed twisted density of -1.1 degrees/bp. **e-f.** Same as C and D but for the left-handed DNA turbine, highlighting the twist density of +0.69 degrees/bp. **g.** Schematic of a right-handed DNA turbine with its load, a 300-nm long DNA bundle with the middle 220-nm reinforced with 16 DNA helices instead of 6 helices, and a 900 nm looped leash docked onto a solid-state silicon nitride nanopore. **h.** SNUPI-simulated structure of the right-handed DNA turbine with a DNA bundle attached as a load (leash excluded). **i.** Negatively stained transmission electron micrograph of a typical right-handed DNA turbine with the load.

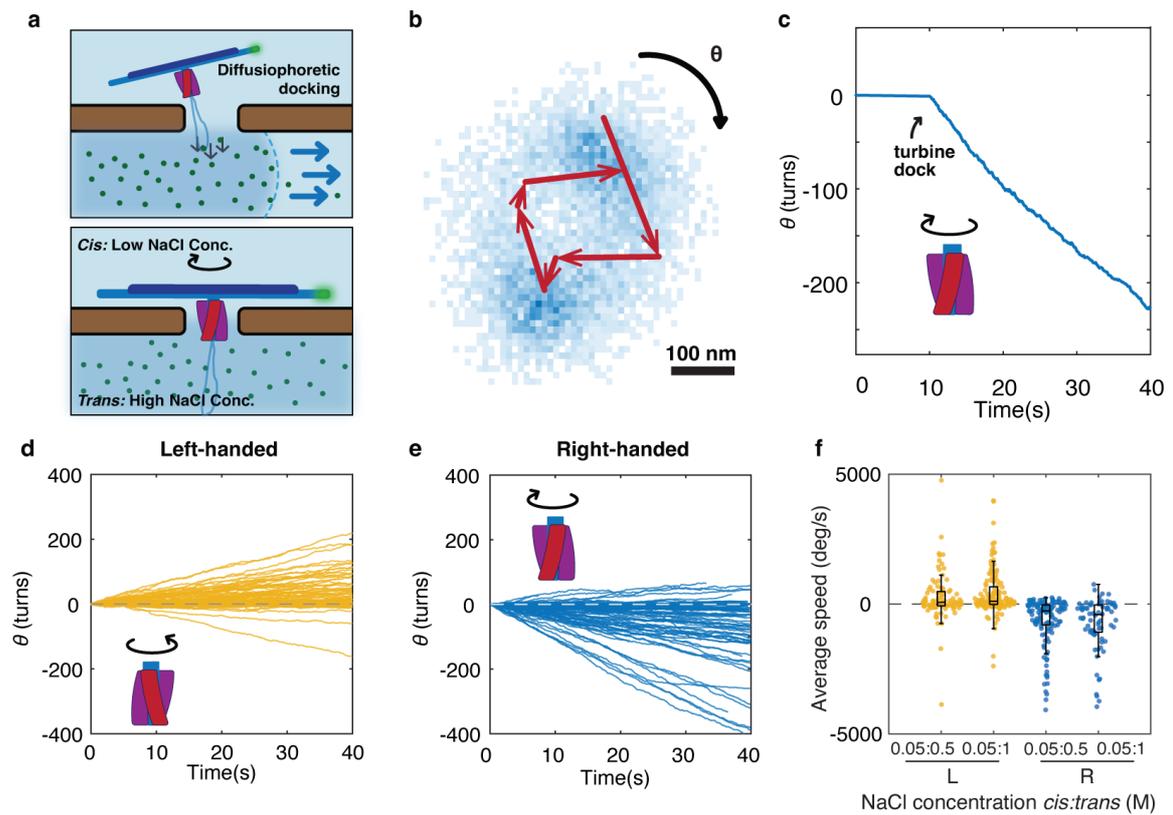

**Figure 2. Sustained unidirectional rotation of DNA origami turbines in a salt gradient.**
**a.** Schematic of a DNA turbine docking onto a nanopore by diffusiophoresis. **b.** Typical heatmap (blue pixels) of obtained centers of fluorophores at the tip of DNA bundle from single-particle localizations from 8000 frames (see Methods). Example trajectory of 6 subsequent positions of the labelled tip (red) which shows clear directional rotation. **c.** Typical cumulative angle versus time for a right-handed turbine in 50 mM:1M NaCl, showing a sustained rotation of hundreds of turns. **d, e.** Cumulative angle versus time of left-handed (d) and right-handed (e) turbines for a NaCl concentration gradient of 50 mM:1 M (n=77 and 151, respectively). **f.** Average rotation speed of left- and right-handed turbines in transmembrane NaCl concentration gradients of 50 mM:500 mM and 50mM:1 M. (n=98, 141, 124, and 74, respectively). In all box plots: center line, median; box limits, upper and lower quartiles; whiskers, 1.5x interquartile range.

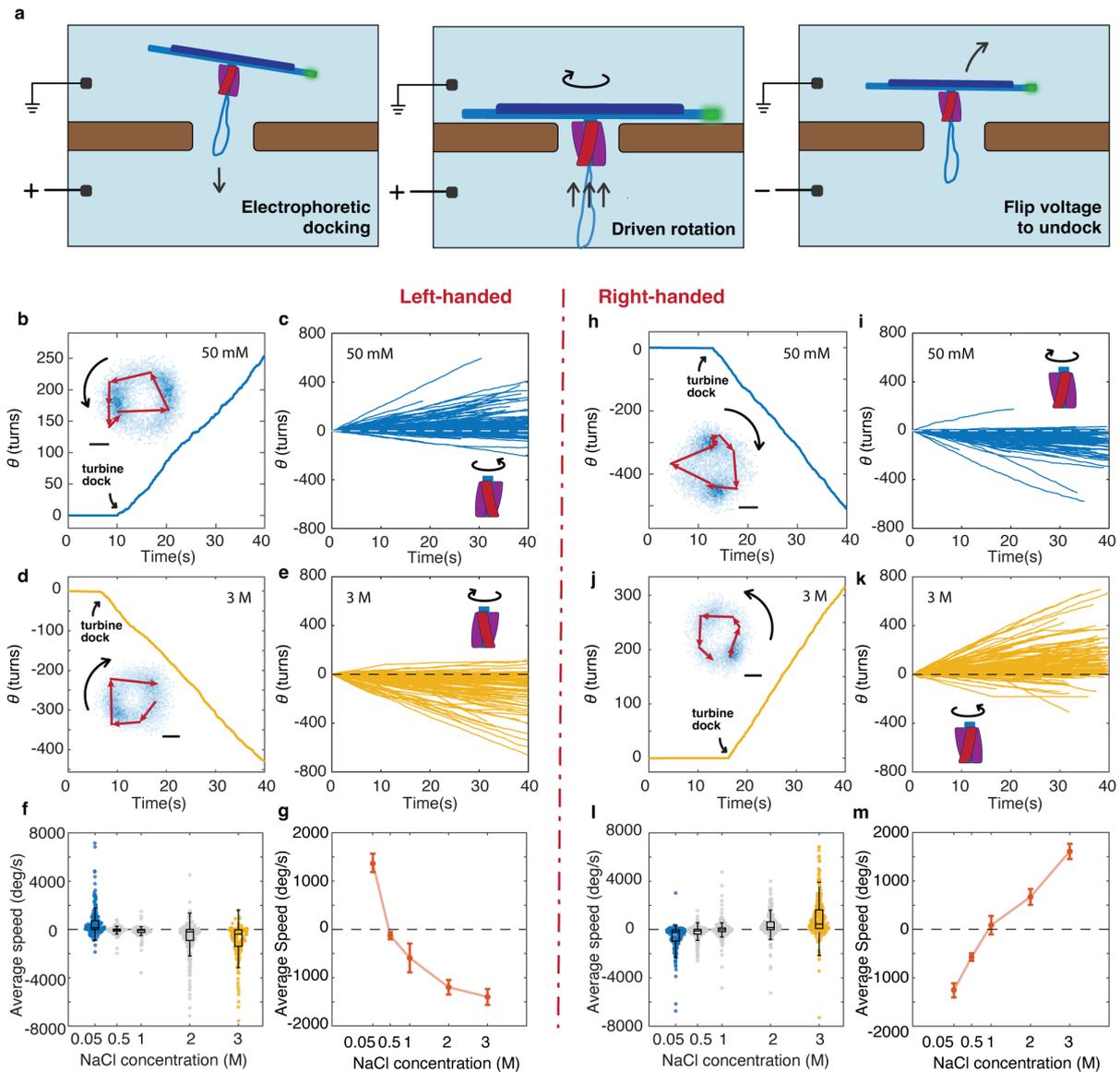

**Figure 3. Sustained unidirectional rotation of DNA origami turbines in an applied field.**
**a.** Schematic of a DNA turbine docking and undocking onto a nanopore by applying a transmembrane voltage. **b.** Typical cumulative angle versus time for a left-handed turbine for a 100-mV bias voltage in 50 mM NaCl, showing a sustained rotation over hundreds of turns. Inset: corresponding heatmap (blue pixels) of single-particle localizations for the tip of DNA bundle with an example trajectory of the labelled tip overlaid. **c.** Cumulative angular-displacement curves for left-handed turbine driven DNA bundles as in panel B but for n = 210 turbines. **d, e.** Same as (b) and (c) but in 3M NaCl electrolyte (n = 159). **f.** Average rotation speed for left-handed turbines in NaCl concentration of 50 mM, 500 mM, 1M, 2M, 3M (n = 198, 77, 86, 252, and 150 respectively).. **g.** Mean rotary speed of left-handed turbines for various buffer salt concentrations (see Supplementary Figure S8). Error bars are standard errors of the mean. **h-m.** Same as (b-g) but for right-handed DNA turbines ($n_i$ = 174, $n_k$ = 298, $n_l$ = 116, 252, 164, 200, and 260. respectively). In all box plots: center line, median; box limits, upper and lower quartiles; whiskers, 1.5x interquartile range.

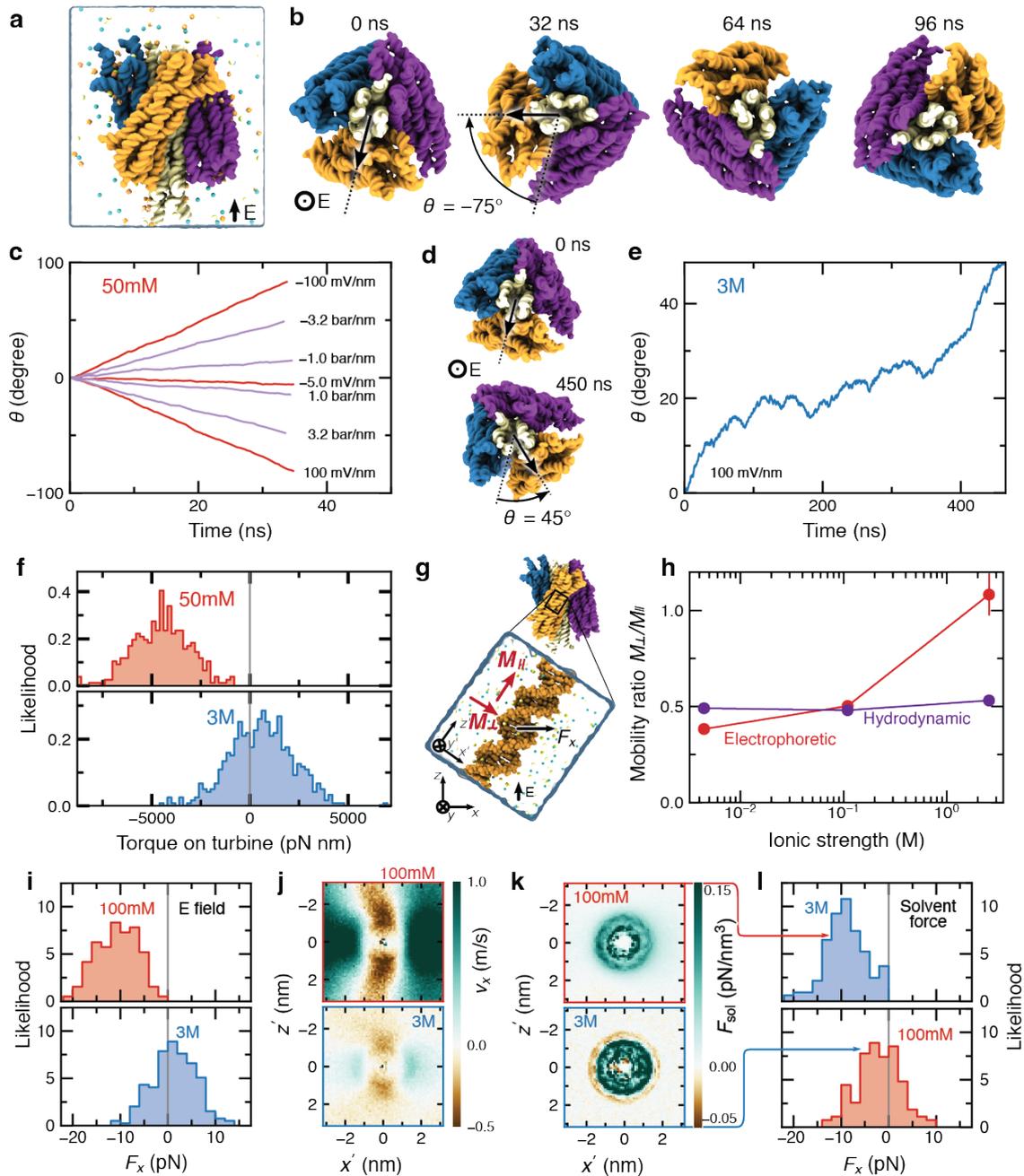

**Figure 4. All-atom MD simulation of a DNA turbine rotation. a.** 4,322,088-atom model of a DNA origami turbine that is depicted using a molecular surface representation (white shaft, multi-coloured blades), solvent shown as a semi-transparent surface, and ions (50 mM NaCl, 10 mM $Mg^{2+}$) shown explicitly. **b.** Electric field-driven rotation of the turbine (50 mM NaCl). A 100 mV/nm field was applied out of the page while restraints prevented the drift and tilting of the turbine. **c.** Rotation angle of the turbine due to an applied field or pressure gradient. **d, e.** Same as panels (b) and (c), but for simulations done at 3 M NaCl. A reversal of the rotation direction is observed. **f.** Torque on the turbine measured by restraining its spin angle. **g.** Single DNA helix as a minimal model for a turbine blade. A field or pressure gradient is applied along the +z direction **h.** Ratio of the electrophoretic and hydrodynamic mobilities for motion perpendicular and parallel to the DNA helical axis observed in simulations of a single DNA helix. Ionic strength is calculated from the average molality observed at distances beyond 6 nm from the helical axis. **i.** Force on DNA orthogonal to the electric field measured from the single helix simulations. **j.** In-plane solvent flow along the $x$ axis in the simulations under an applied electric field. Heat maps depict the average along the helical axis of the DNA. The coordinate system is defined in panel (g).

**k.** Solvent forces extracted from MD simulation under an applied electric field. **l.** Force on DNA orthogonal to the applied force axis when, as a control, the solvent forces extracted from the low and high salt systems are applied to high and low salt systems, respectively. The forces were seen to reverse compared to those in panel (i), proving the causal role of the ion distribution.

## Methods

**Nanopore array fabrication**

Nanopore arrays are fabricated as reported before[27]. In brief, a 100-nm thick layer of poly (methyl methacrylate) (PMMA) electron-sensitive resist (molecular weight 950k, 3% dissolved in anisole, MicroChem Corp) was spin-coated on 20 nm free-standing silicon nitride membranes supported by silicon. Subsequently, the resist was exposed and patterned by an electron beam pattern generator (EBPG5200, Raith) with 100 keV electron beams. The pattern is developed in a mixture of methyl-isobutyl-ketone (MIBK) and isopropanol (IPA) with a ratio of 1:3 for 1 min, then stopped in IPA for 30 sec. The exposed substrates were then etched using RIE with fluoroform and argon (200 s, 50 W, 50 sccm of $CHF_3$, 25 sccm of Ar, 10 µbar, Sentech Plasma System SI 200). Finally, the resist was removed in oxygen plasma for 1 min (200 $cm^3$/min $O_2$, 100 W, PVA Tepla 300) followed by an acetone bath for 5 min.

**Design, folding, and purification of DNA origami structures**

All structures were designed using cadnano v0.2[28]. For the cryo-EM reconstruction of the turbine part, all structures were designed with a compact beam on top of each turbine structure (Fig. S6 and S7) and designed only using a 7560-bases long scaffold. The folding reaction mixtures contained a final scaffold concentration of 50 nM and oligonucleotide strands (IDT) of 500 nM. The folding reaction buffer contained 5 mM Tris, 1 mM EDTA, 5 mM NaCl and 20 mM $MgCl_2$. The folding solutions were thermally annealed using TETRAD (MJ Research, now Biorad) thermal cycling devices. The reactions were left at 65 °C for 15 minutes and then subsequently subjected to a thermal annealing ramp from 60 °C to 20 °C (1 °C/hour). The folded structures were purified from excess oligonucleotides by physical extraction from agarose gels and stored at room temperature until further usage. The list of oligos can be found in the Supplementary Data.

The turbine structure with a long DNA bundle as load were designed using a scaffold of 8064-bases and a scaffold of 9072-bases length. the folding reaction mixtures contained a final scaffold concentration of 10 nM plus oligonucleotide strands (IDT) of 100 nM each. The folding reaction buffer contained 5 mM Tirs, 1 mM EDTA, 5 mM NaCl and 15 mM $MgCl_2$ for the left-handed and right-handed versions or 20 mM $MgCl_2$ for the achiral version of the turbine. The folding reaction mixtures were thermally annealed using TETRAD (MJ Research, now Biorad) thermal cycling devices. The reactions were left at 65 °C for 15 minutes and then subjected to a thermal annealing ramp from 60 °C to 20 °C (1 °C/hour). The folded structures were purified from excess oligonucleotides by PEG precipitation and stored at room temperature until further usage. Details of all the procedures can be found in [29].

**Cryogenic electron microscopy sample preparation, image acquisition and processing**

Grid preparation, image acquisition, and data processing were largely performed as reported previously [30]. The sample was applied to a glow-discharged C-Flat 1.2/1.3 4C Thick grid (Protochips) and vitrified using a Vitrobot Mark IV (FEI, now Thermo Scientific) at a temperature of 22 °C, a humidity of 100 %, 0 s wait time, 2 s blot time, -1 blot force (arbitrary device units), and 0 s drain time. 3,427 and 5,997 micrograph movies with 10 frames were collected for the right-handed and left-handed versions, respectively, at a magnified pixel size of 2.28 Å and an accumulated dose of ~60 e/Å$^2$ using the EPU software and a Falcon 3 detector (FEI, now Thermo Scientific) on a Cs-corrected (CEOS) 300-kV Titan Krios electron microscope (FEI, now Thermo Scientific). For the left-handed version, acquisition with a stage tilt of 20° was used to reduce the orientation bias of the particles.

Motion correction and contrast transfer function (CTF) estimation of the micrographs were performed using the implementation in Relion 4.0 beta[31,32] and CTFFIND4, respectively [33]. Particles were auto-picked using TOPAZ [34] and subjected to a selection process consisting of multiple rounds of 2D and 3D classification in Relion to remove falsely picked particles and damaged particles. Using an ab-initio initial model, a refined 3D map was reconstructed from 97,054 and 71,992 particles for the right-handed and left-handed versions, respectively, followed by per-particle-motion correction and dose weighting and 3D refinement (Fig. S6 and S7). For a focused reconstruction of the turbine, multibody (MB) refinement[35] was performed. The consensus map was divided into two parts containing the lever and the turbine using the eraser tool in UCSF Chimera[36] and low pass filtered soft masks of the respective regions were created in Relion (Fig. S8 and S9). After MB refinement, a set of particles with the subtracted signal of the lever arm was calculated and subjected to another round of 3D refinement. The final maps were masked, sharpened, and low pass filtered using the estimated resolution based on the 0.143 FSC criterium. Atomic models were constructed using a cascaded relaxation protocol as described previously [30] (Fig. S11).

The dimensions of the turbines were measured in Fiji[37] using orthographic projections of the maps created with ChimeraX[38]. For the twist measurement of the turbine versions, slices from the well-resolved central parts were extracted from the cryo-EM density maps using atomic model fits at base pair positions which are on the same plane in the design with a spacing of 33 bp and 34 bp for the right- and the left-handed version, respectively (Figure S10). For each version, the slices were fitted into each other based on maximum overlay using ChimeraX[38] to determine the rotation angle. From the twist density, the diameter, and the length of the helices, the outer blade angle with respect to the helical axes was calculated.

**Single-particle fluorescence imaging**

Solid-state nanopore chips were oxygen-plasma cleaned before all the fluorescence experiments (100 W for 1 min, Plasma Prep III, SPI Supplies). Coverslips (VWR, No. 1.5) were cleaned by ultrasonication sequentially in acetone, isopropanol, water, 1 M KOH solution, and deionised water (DI, Milli-Q) for 30 min each. The cleaned coverslips were then blow-dried thoroughly with compressed nitrogen. The nanopore chip was glued into the PDMS (SYLGARD™ 184 Silicone Elastomer) flow cell using a 2-component silicone rubber (Ecoflex 5, Smooth-ON), then the PDMS flow cell was bonded to the cleaned coverslip after oxygen-plasma treatment (50 W, 50 mbar for 30 s) and post-bake under 120 °C for 30 min. After assembly, the whole device was again treated with oxygen plasma (50 W, 50 mbar) for 4 min before embedding a pair of Ag/AgCl electrodes, one in each side of the reservoir and flushing in DI water to wet the channels. This is essential for increasing the hydrophilicity of the membrane and ensuring a negatively charged silicon nitride surface. The PDMS-nanopore devices were always assembled shortly before each experiment and never re-used.

The nanopore chip was then imaged using an epifluorescence microscope with a 60x water immersion objective (Olympus UplanSApo 60x, NA1.20) and a fast sCMOS camera (Prime BSI, Teledmy Photometrics). The camera field of view was reduced as needed to achieve high frame rates (typically around 200 pixels x 200 pixels). To image Cy3 labelled DNA turbines, a 561 nm laser (Stradus®, Vortran Laser Technology) was used to excite the fluorophores. The typical exposure time of the experiments was 5 ms, which led to a frame rate of around 190-200 fps. To simplify the data analysis, a fixed frame rate value (200 fps) is used. Before imaging, the imaging buffer (50 mM Tris-HCl pH 7.5; 50 mM NaCl unless otherwise stated, 5 mM $MgCl_2$; 1 mM DTT, 5% (w/v) D-dextrose, 2 mM Trolox, 40 μg ml$^{-1}$ glucose oxidase, 17 μg ml$^{-1}$ catalase; 0.05% TWEEN20) was added into the two reservoirs on each side of the silicon nitride membrane.

**Driving DNA turbines using transmembrane ion gradients**

An imaging buffer with the same salt concentration (50 mM NaCl) was flushed into the flow cell first, with DNA turbines on the *cis* side of the membrane. Subsequently, an imaging buffer containing a higher NaCl concentration was flushed into the *trans* side of the membrane. With single-particle fluorescence microscopy, the docking and rotation of the DNA turbines could be observed and recorded. To release the turbines from the nanopore, we either inserted a pair of temporary electrodes into the inlet and outlet of the flow channels and released the turbines electrically or we flushed in the same (lower concentration) buffer as the *cis* side. Because of the photobleaching and accumulation of the DNA turbines near nanopore arrays, we chose 40 seconds as a typical observation duration. Examples of longer recordings are shown in Supplementary Figure S23

**Driving DNA turbines using transmembrane voltages**

Different from the salt-gradient-driven mode, a pair of electrodes were embedded into the flow cells. We used a custom-built circuit to apply voltages[39]. The output voltage was controlled by a custom LabVIEW program. The electrodes embedded in the two reservoirs were connected to the circuit. The DNA origami turbines were added into the electrically grounded side (*cis* side) of the flow cell with a typical concentration of 1 pM. After applying the voltage, the DNA turbines were docked onto the nanopores under a 100-mV bias voltage (unless otherwise stated) across the membrane. The turbines could be easily released from the nanopore array by flipping the voltage polarity and then setting it to 0 mV for several seconds to allow the imaged turbines to diffuse away from the capture region. To avoid overcrowding of DNA turbines near the nanopore array, which increased the fluorescence background fluctuation, we typically imaged the turbines before the array was fully filled, and subsequently released them from the nanopore. Then a new group of turbines could be captured and docked again by applying a positive bias.

**Fluorescence microscopy data analysis**

For image processing, first, a single-molecule localisation was done by using Fiji (ImageJ[37]) with the ThunderStorm plugin[40] for all frames in the acquired image sequences. A wavelet filter (B-spline) and an integrated Gaussian method were used for the localisations. Then the results were filtered based on their quality (uncertainty < 50nm) and the local density (filter of 15 particles in every 50 nm among all localised data points in the sequence) to rule out free diffusing (non-captured) turbines. Next, the single-molecule localisation results were analysed using a custom MATLAB script (see Code Availability). In brief, all coordinates of localised particle positions were clustered based on their Euclidean distance for each turbine. When localised particle positions were deduced in a movie, a circle was fit to the data to obtain the centre and a radius, which subsequently was used for calculating the angular position of the fluorophores in each frame. Next, we determined if the fluorophores occupy spatial states that can be fitted to a circular path. We did this by comparing the point density of the coordinates within an annulus around the fitted circular perimeter (±1 nm) with the density of points around the centre (with a radius *r*, so that the area of this central circle equals that of the ring region). If the point density within the annulus was higher, then this data group would be kept in the statistics, else it would be considered an invalid trajectory and discarded. Finally, the script calculated all necessary motion properties of the turbine, including its cumulative angular displacements, MSD, angular velocity, and torque. The angular velocity $\omega_d$ was determined by fitting $MSD = \omega_d^2 t^2 + 2D_r t$ to the MSD curve of each turbine, where $t$ is the lag time and $D_r$ is the rotational diffusion coefficient (also as a fitting parameter). The estimation of the torque is discussed in Supplementary Note 1.

**Molecular dynamics simulations**

All MD simulations were performed using the NAMD program[41], CHARMM36 parameters for DNA, water and ions[42] with CUFIX [43] corrections, periodic boundary conditions, and the TIP3P model of water[44]. The long-range electrostatic interactions were computed using the particle-mesh Ewald scheme over a grid with 1 Å-spacing[45]. Van der Waals and short-range electrostatic forces were evaluated using the 10–12 Å smooth cutoff scheme. Hydrogen mass repartitioning[46], and the SHAKE [47] and SETTLE [48] algorithms were used, enabling a 4 fs integration time step. The full electrostatics were calculated every two-time step. Except where specified, a Langevin thermostat with a 0.1 ps$^{-1}$ damping coefficient maintained a temperature of 295 K in all simulations. Coordinates were recorded every 2500 steps.

Atomistic models of the entire turbine were assembled from the cadnano[28] design file using a custom mrDNA script[49]. Neutralising plus 10 mM Mg$^{2+}$ hexahydrate were placed adjacent to the DNA according to a previously described protocol[43]. Water and monovalent ions were added to the system using the solvate and autoionise plugins for VMD[50], with the solvent box cut to form a hexagonal prism. For each salt condition, a 4 ns simulation was performed with a Nosé–Hoover Langevin piston barostat [51,52] set to maintain a target pressure of 1 bar, allowing the equilibrium volume of the system to be determined. The resulting system dimensions were used in constant volume simulations to equilibrate the turbine with harmonic position restraints holding the phosphorus atoms to their initial coordinates during the first 7 ns of the simulation ($k_{spring}$ = 1 kcal mol$^{-1}$ Å$^{-2}$ for $t$ < 5ns; 0.1 for 5ns < t < 7ns). After 30 ns of equilibration for the 50 mM and 75 ns for the 3M system, a snapshot of the configuration was used to initialise subsequent simulations with either an electric field or pressure gradient applied to drive the turbine. Additional equilibration was performed for the 50 mM NaCl system for another 128 ns to initialise the "Alternate conf." system, see SI Fig. S24.

The conformation of the turbine at the end of equilibration was used to determine the rest positions of several harmonic collective variables (colvar) [53] potentials, including a spring restraining the centre of mass (CoM) of every third phosphorus atom ($k_{spring}$ = 500 kcal mol$^{-1}$ Å$^{-2}$); a spring restraining the RMSD of those phosphorus atoms with respect to the post-equilibration configuration ($k_{spring}$ = 1000 kcal mol$^{-1}$ Å$^{-2}$; resting RMSD = 0), after optimal rigid body transformations so that the potential does not apply a net torque or force; and a pair of CoM harmonic restraints applied to 16-bp-long sections of the central six-helix bundle ($k_{spring}$ = 50 kcal mol$^{-1}$ Å$^{-2}$), placed near either the end of the shaft to prevent the turbine from tilting. With these colvars preventing translation, conformational fluctuations, or tilting of the turbine, an electric field was applied by placing a constant force on each atom with a magnitude proportional to the charge of the atom. Similarly, a pressure gradient was achieved by placing a small force on every water molecule of the system. Finally, in simulations where the torque was measured, an additional spin angle colvar ($k_{spring}$ = 100 kcal mol$^{-1}$ degree$^{-2}$) prevented rotation of the turbine and reported the torque.

Simulation systems were prepared to study the forces on and flows around a DNA helix mimicking the DNA in the turbine blade. The 21-bp helix was made effectively infinite by connecting the ends of each strand across the periodic boundary. Solvent (neutralising Na$^+$, 100 mM and 3 M NaCl; no Mg$^{2+}$) was added around the helix. Systems were equilibrated for 5-50 ns with the DNA phosphorus atoms harmonically restrained ($k_{spring}$ = 0.2 kcal mol$^{-1}$ Å$^{-2}$). Mobility measurements were performed using a field of 5 mV/nm or a hydrostatic pressure of ~1.3 bar/nm parallel or transverse to the helical axis, with each condition employing four replicate simulations lasting a total of 400 (neutralising Na$^+$) to 4000 ns (3 M). Except where otherwise specified, a 100 mV/nm electric field was applied to the system at a 35° angle with respect to the DNA while a CoM colvar restrained the DNA ($k_{spring}$ = 500 kcal mol$^{-1}$ Å$^{-2}$), an RMSD colvar retained an idealised DNA configuration ($k_{spring}$ = 100 kcal mol$^{-1}$ Å$^{-2}$), and a spin angle colvar ($k_{spring}$ = 100 kcal mol$^{-1}$ degree$^{-2}$) prevented rotation of the DNA and reported on the torque. ight replicate systems were employed during simulations lasting a total of 1040 ns (100 mM) or 2055 ns (3 M). The flow and concentration of ions and water oxygen atoms were analysed by binning the system into ~1 Å voxels, counting the flux through and concentration in each voxel using a centred finite difference approximation for the flux. The difference in concentration between sodium and chloride ions provided the net local charge density of the fluid around the DNA in each case. Multiplying that charge by the electric field provided the solvent force. In subsequent simulations, the 3D map of the solvent force was used to apply a position-dependent force to each water oxygen atom using the TclBC feature of NAMD and employing the approximation that the density of water oxygen atoms is uniformly 33 per nm$^3$. Again, eight replicate systems were employed for simulations lasting a total of 480 and 570 ns for 100 mM and 3 M conditions, respectively.

**Statistics & Reproducibility**

No statistical method was used to predetermine the sample size.

**Data Availability**

The electron density maps of the left- and right-handed turbine are available in the electron microscopy data bank (EMDB) as entries EMD-17600 and EMD-17606, respectively. Fluorescence microscopy and nanopore experimental data are available at https://doi.org/10.5281/zenodo.8091178. Simulation trajectory data is available at https://doi.org/10.13012/B2IDB-3458097_V1.

**Code Availability**

The corresponding MATLAB scripts for data processing and producing the final figures are available at https://doi.org/10.5281/zenodo.8091178. Scripts related to simulation setup and analysis are available at https://doi.org/10.13012/B2IDB-3458097_V1.

**Methods-only References**

**This PDF file includes:**

Supplementary Text
Supplementary Figures. S1 to S26
Supplementary Movie Captions

**Other Supplementary Materials for this manuscript include the following:**

Supplementary Movie S1-S2: Simulated DNA turbine driven by voltage.
Supplementary Table S1: DNA origami sequences for the turbines.

# Supplementary Text

## 1. Estimate the torque of the turbine from MSD curves.

We estimate the torque that the turbine exerts as follows. We adopt the following assumptions about the system: 1) We assume that the DNA bundle behaves as a passive load; 2) we simplify the bundle to a stiff cylindrical-shaped rod with dimensions of a 250 nm length and 12 nm diameter, i.e. considering only slightly longer than the middle reinforced part of the used DNA bundle to contribute to the load; and 3) we neglect any surface effects resulting from the close proximity of the membrane to the DNA bundle.

For purely rotational diffusion about a single axis, the mean-square angular deviation in time $t$ is

$$\langle\theta^2\rangle = 2D_r t, \tag{1}$$

where $\theta$ is the measured rotation angle (in rad), $D_r$ is the rotational diffusion coefficient (in rad$^2$/s). Additionally, ignoring diffusion, the angular drift velocity $\omega_d = (d\theta/dt)_{drift}$ in response to an external torque $\Gamma_\theta$ results in

$$\langle\theta^2\rangle = \omega_d^2 t^2,$$

with

$$\omega_d = \frac{\Gamma_\theta}{f_r}, \tag{2}$$

where $f_r$ is the frictional drag coefficient. The relationship between the rotational diffusion coefficient and the rotational frictional drag coefficient is given by Einstein–Smoluchowski relation:

$$D_r = \frac{k_B T}{f_r}, \tag{3}$$

where $k_B$ is the Boltzmann constant and $T$ is temperature. These relationships are in complete analogy to translational diffusion. The hydrodynamic drag of the DNA rod is estimated by treating it as a cylinder of length 250 nm and 6 nm radius rotating about its middle point along the long axis, for which $f_r$ can be estimated as [1]

$$f_r = \frac{\pi \eta L^3}{3(\ln p + \delta_T)}, \tag{4}$$

where $\eta$ is the solvent viscosity, $L$ is the cylinder length, $\delta_T$ is an end-effect correction, and

$$p = \frac{L}{2R}, \tag{5}$$

where $R$ is the cylinder radius. For our system, we have $\eta = 10^{-3}\,\text{Pa}\cdot\text{s}, L = 250\,\text{nm}, R = 6\,\text{nm}, p = \frac{250}{2\times 6} \approx 20.8$, and we use $\delta_T = -0.616$ as an approximate end-effect correction [1].

Now the contributions to the angular displacement of in response to both diffusion and external torque add linearly and we can fit the model

$$\langle\theta^2\rangle = a_1 t^2 + a_2 t = \omega_d^2 t^2 + 2D_r t, \tag{6}$$

to our data for $\theta(t)$, where $\omega_d$ is the angular drift velocity, $a_1$ and $a_2$ are the two fitting parameters, as

$$a_1 = \left(\frac{\Gamma_\theta}{f_r}\right)^2, \tag{7}$$

From equation (1) we thus obtain the torque

$$\Gamma_\theta = f_r\sqrt{a_1}, \tag{8}$$

This was implemented into a MATLAB routine to estimate torques for every DNA turbine, and the results are shown in Fig. S19.

## 2. Continuum hydrodynamic model for the reversal of the rotational direction

Consider a high-aspect ratio rigid cylindrical DNA rod in a wide nanopore that is orientated at an angle of $\theta$ with respect to the Z axis, as shown in the schematic in Scheme S1.

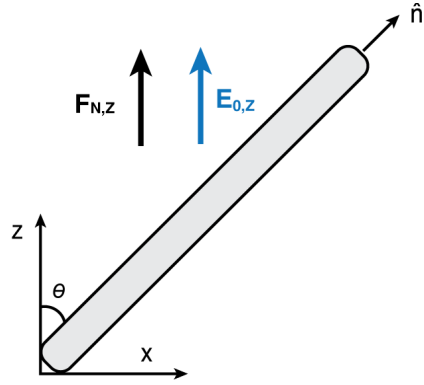

**Scheme S1. A rigid cylindrical DNA rod.**

The net motion of this single cylindrical DNA rod that is held in the applied $E_z$ field in the nanopore is governed by the equation:

$$\boldsymbol{v} = [M_{el,\parallel} \hat{n}\hat{n} + M_{el,\perp}(\boldsymbol{I} - \hat{n}\hat{n})] \cdot \boldsymbol{E_Z} + [M_{h,\parallel} \hat{n}\hat{n} + M_{h,\perp}(\boldsymbol{I} - \hat{n}\hat{n})] \cdot \boldsymbol{F_{N,Z}}, \tag{1}$$

$$v_z = [M_{el,\parallel} \cos^2\theta + M_{el,\perp}\sin^2\theta] \cdot E_Z + [M_{h,\parallel} \cos^2\theta + M_{h,\perp}\sin^2\theta] \cdot F_{N,Z} \tag{2}$$

$$v_x = \sin\theta \cos\theta \left[(M_{el,\parallel} - M_{el,\perp}) E_Z + (M_{h,\parallel} - M_{h,\perp}) F_{N,Z}\right], \tag{3}$$

where $v_z = 0$ because the rod is vertically held in position. Hence it follows that

$$F_{N,Z} = -\left(\frac{M_{el,\parallel}\cos^2\theta + M_{el,\perp}\sin^2\theta}{M_{h,\parallel}\cos^2\theta + M_{h,\perp}\sin^2\theta}\right) \cdot E_Z \tag{4}$$

Combining (3) and (4) yields and expression for $v_x$ as follows

$$v_x = \sin\theta \cos\theta \left[(M_{el,\parallel} - M_{el,\perp}) - (M_{h,\parallel} - M_{h,\perp})\left(\frac{M_{el,\parallel}\cos^2\theta + M_{el,\perp}\sin^2\theta}{M_{h,\parallel}\cos^2\theta + M_{h,\perp}\sin^2\theta}\right)\right] \cdot E_Z \tag{5}$$

This equation (5) can be rewritten as

$$v_x = \frac{1}{2} M_{el,\|} \sin 2\theta \cdot \sigma \cdot E_z, \tag{6}$$

Where

$$\sigma = \frac{\left(\frac{M_{h,\perp}}{M_{h,\|}} - \frac{M_{el,\perp}}{M_{el,\|}}\right)}{\cos^2\theta + \left(\frac{M_{h,\perp}}{M_{h,\|}}\right)\sin^2\theta} \tag{7}$$

Notably, $M_{h,\perp}/M_{h,\|} = 0.5$ for a thin long rod, while $M_{el,\perp}/M_{el,\|}$ depends on the salt concentration and ranges between 0 and 1. Specifically, $M_{el,\perp}/M_{el,\|}$ depends on the Dukhin number $Du = \frac{\sigma\, l_B}{\kappa^2 b} = \frac{\sigma}{4\pi C_0 b}$, where σ is the surface charge density, 1/κ is the Debye length, and $l_B$ is Bjerrum length , $C_0$ is the salt concentration, and $b$ is the DNA radius. For DNA, $Du = \frac{0.2}{[C_0]}$. Using values from Ref. 2, we can estimate $M_{el,\perp}/M_{el,\|}$ for various salt concentrations as follows:

| $Du$ | $C_0$ | $\frac{M_{el,\perp}}{M_{el,\|}}$ |
| --- | --- | --- |
| 6.43 | 30 mM | 0.14 |
| 0.569 | 350 mM | 0.44 |
| 0.0645 | 3 M | 0.84 |

Based on these numbers, we expect σ to change sign somewhere between 500 mM and 1 M, which indeed is the salt range where the turbine experiments show a sign reversal of the rotations.

An alternative calculation that is similar to the MD simulations is to add a force in the x direction to equation (1) and demand $v_x = 0$. Equation (2) & (3) will then be modified as

$$v_z = [M_{el,\|}\cos^2\theta + M_{el,\perp}\sin^2\theta] \cdot E_Z + [M_{h,\|}\cos^2\theta + M_{h,\perp}\sin^2\theta] \cdot F_{N,Z} \\ + [(M_{h,\|} - M_{h,\perp})\sin\theta\cos\theta] F_x \tag{8}$$

$$v_x = [(M_{el,\|} - M_{el,\perp})\sin\theta\cos\theta] E_Z + [(M_{h,\|} - M_{h,\perp})\sin\theta\cos\theta] F_{N,Z} \\ + [M_{h,\|}\sin^2\theta + M_{h,\perp}\cos^2\theta] F_x \tag{9}$$

Setting $v_z = 0$ and $v_x = 0$, we can eliminate the $F_{N,z}$ between (8) and (9), and solve for $F_x$ as a function of $E_z$. These yields:

$$F_x = -\frac{M_{el,\parallel}}{M_{h,\perp}} \frac{1}{2} \sin 2\theta \left( \frac{M_{h,\perp}}{M_{h,\parallel}} - \frac{M_{el,\perp}}{M_{el,\parallel}} \right) \tag{10}$$

This shows that the sign of $F_x$ is controlled by the same competition between the mobility (anisotropy) ratios. Moreover, we observe that $F_x < 0$ at lower salt concentration and $F_x > 0$ at higher salt concentration (see discussion above), in agreement with the MD simulations results in Fig 4i of the main text.

**Supplementary Figures**

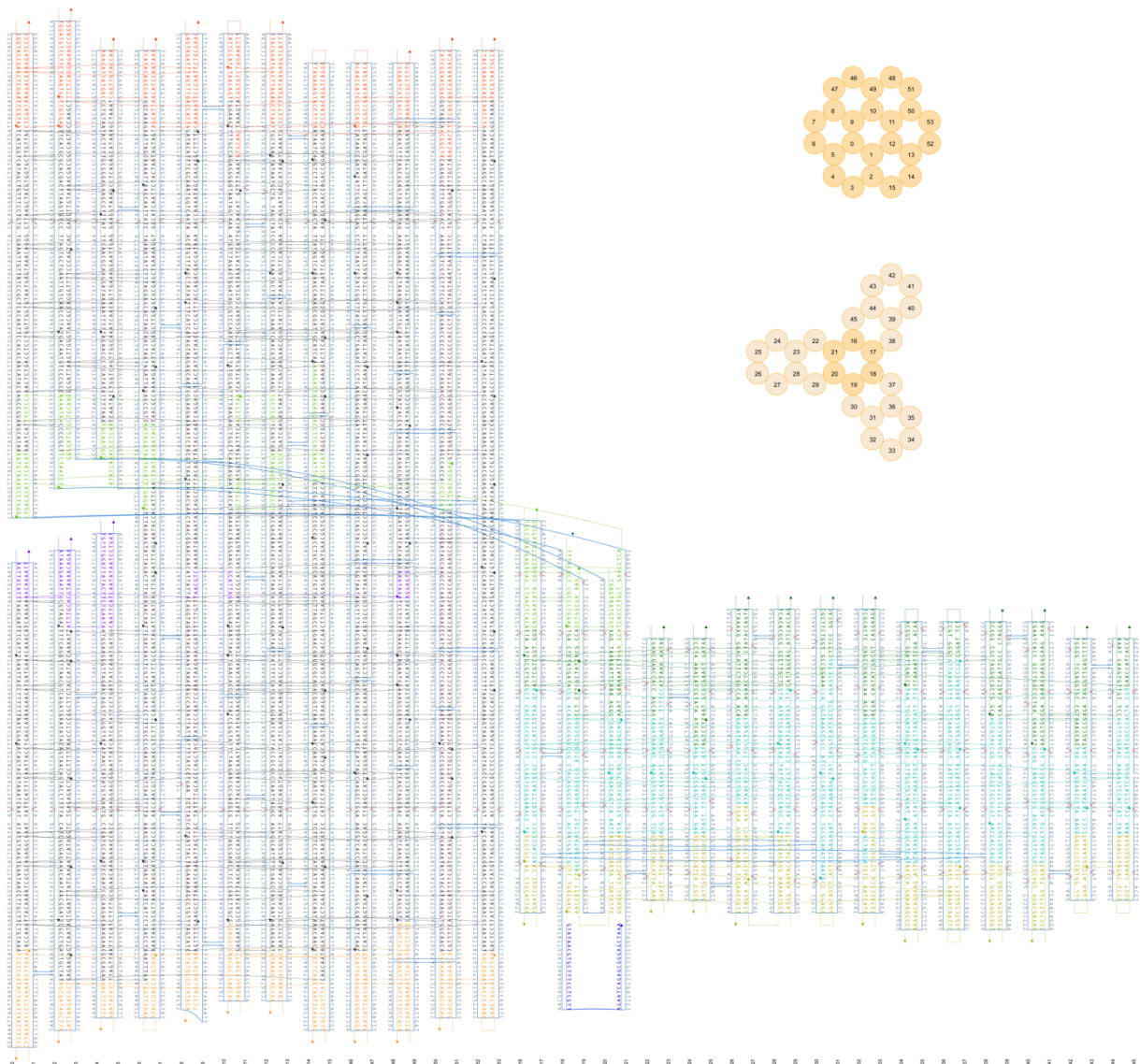

**Figure S1. Design diagram of the left-handed turbine.**
This design was created with caDNAno *(28)*.

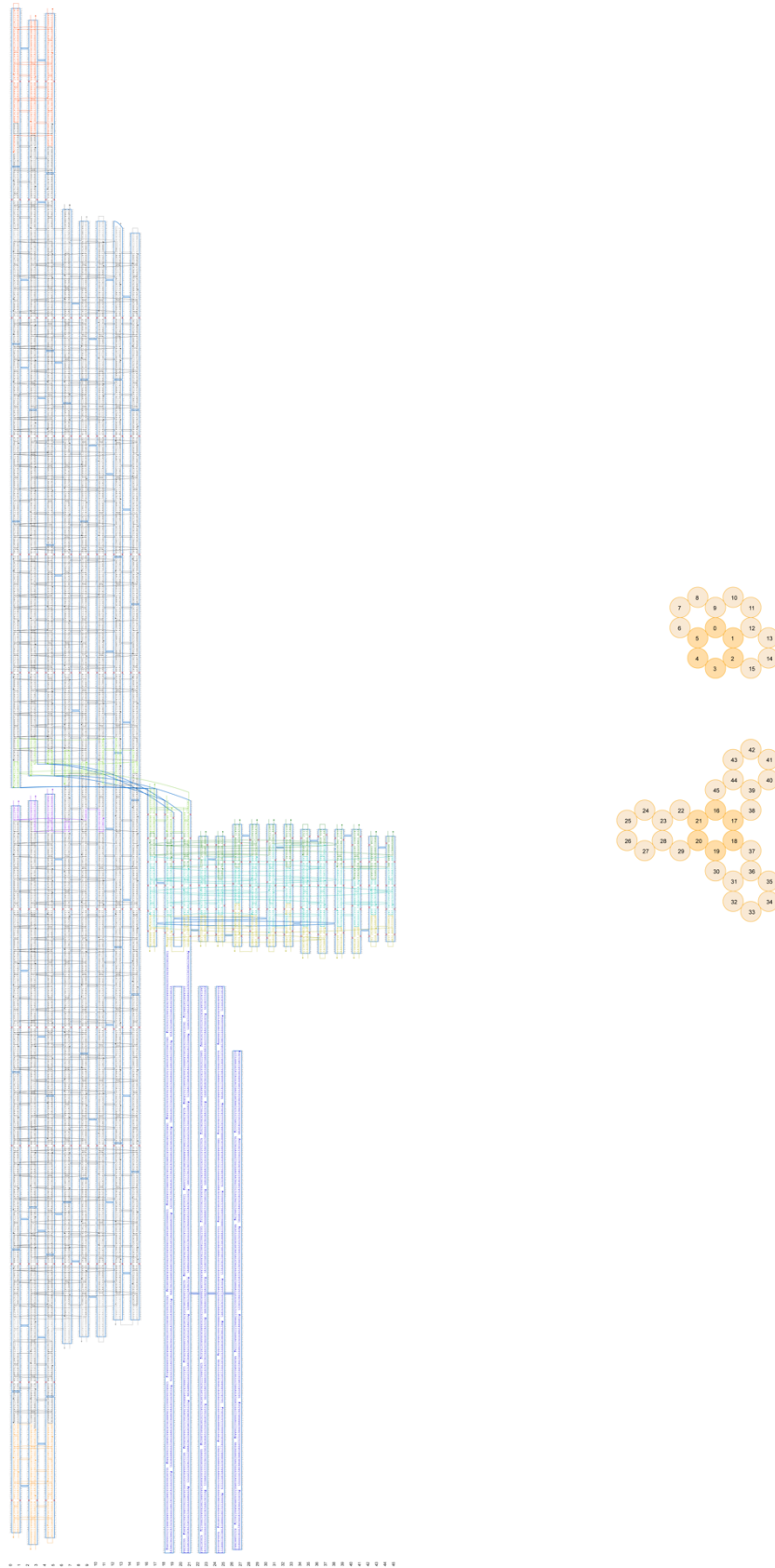

**Figure S2. Design diagram of the left-handed turbine including load.**
This design was created with caDNAno *(28)*.

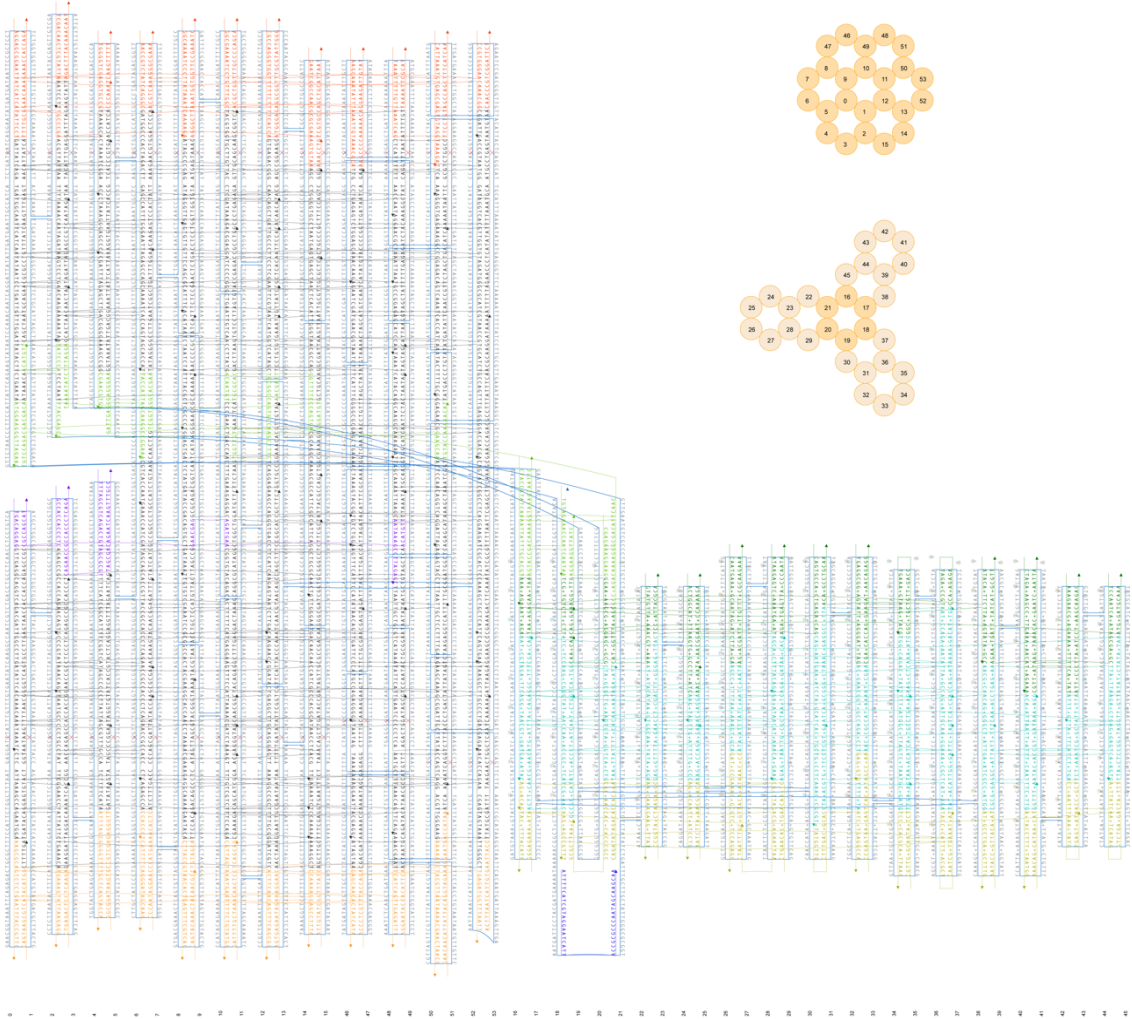

**Figure S3. Design diagram of the right-handed turbine.**
This design was created with caDNAno *(28)*.

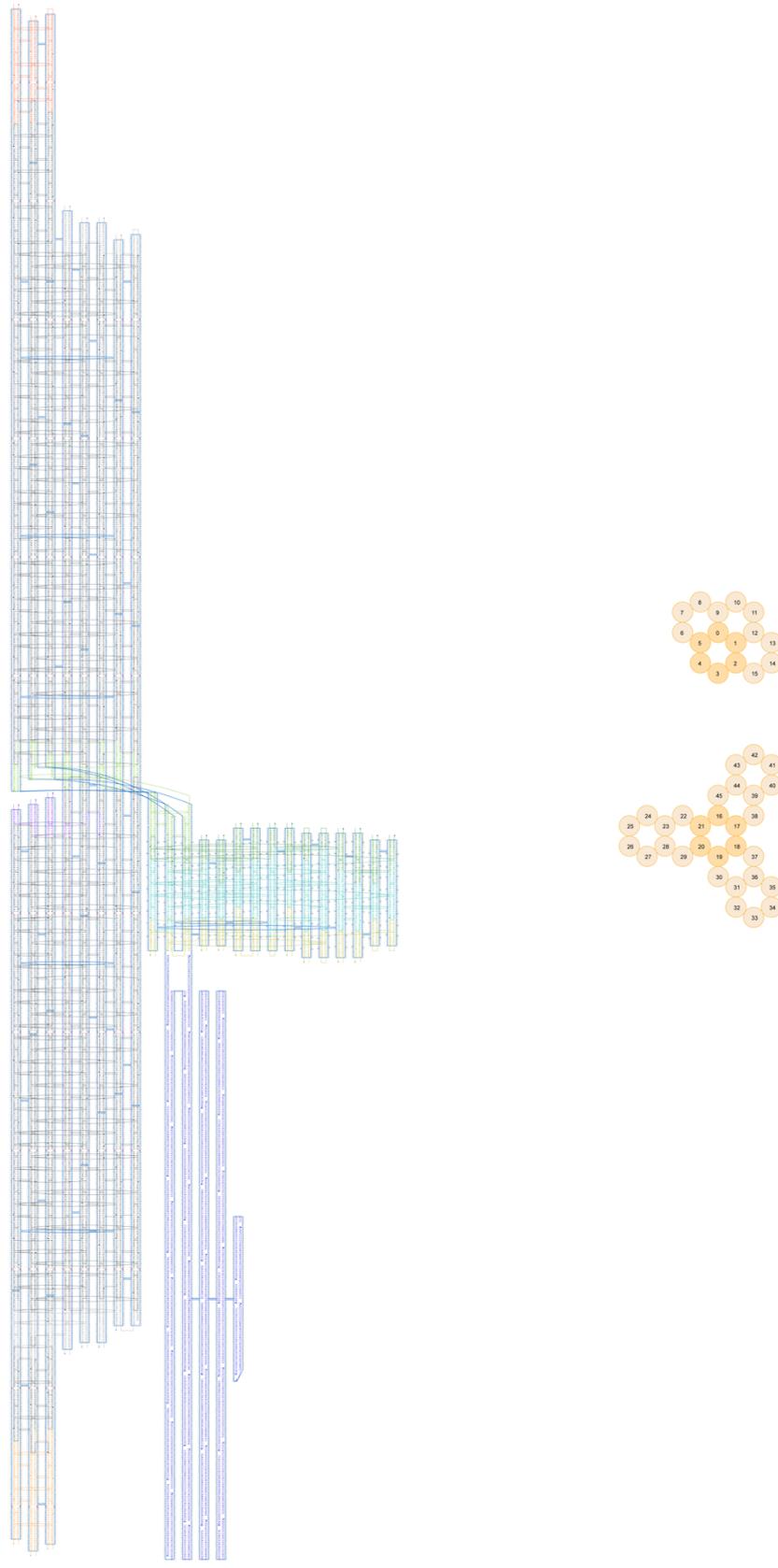

**Figure S4. Design diagram of the right-handed turbine including load.**
This design was created with caDNAno *(28)*.

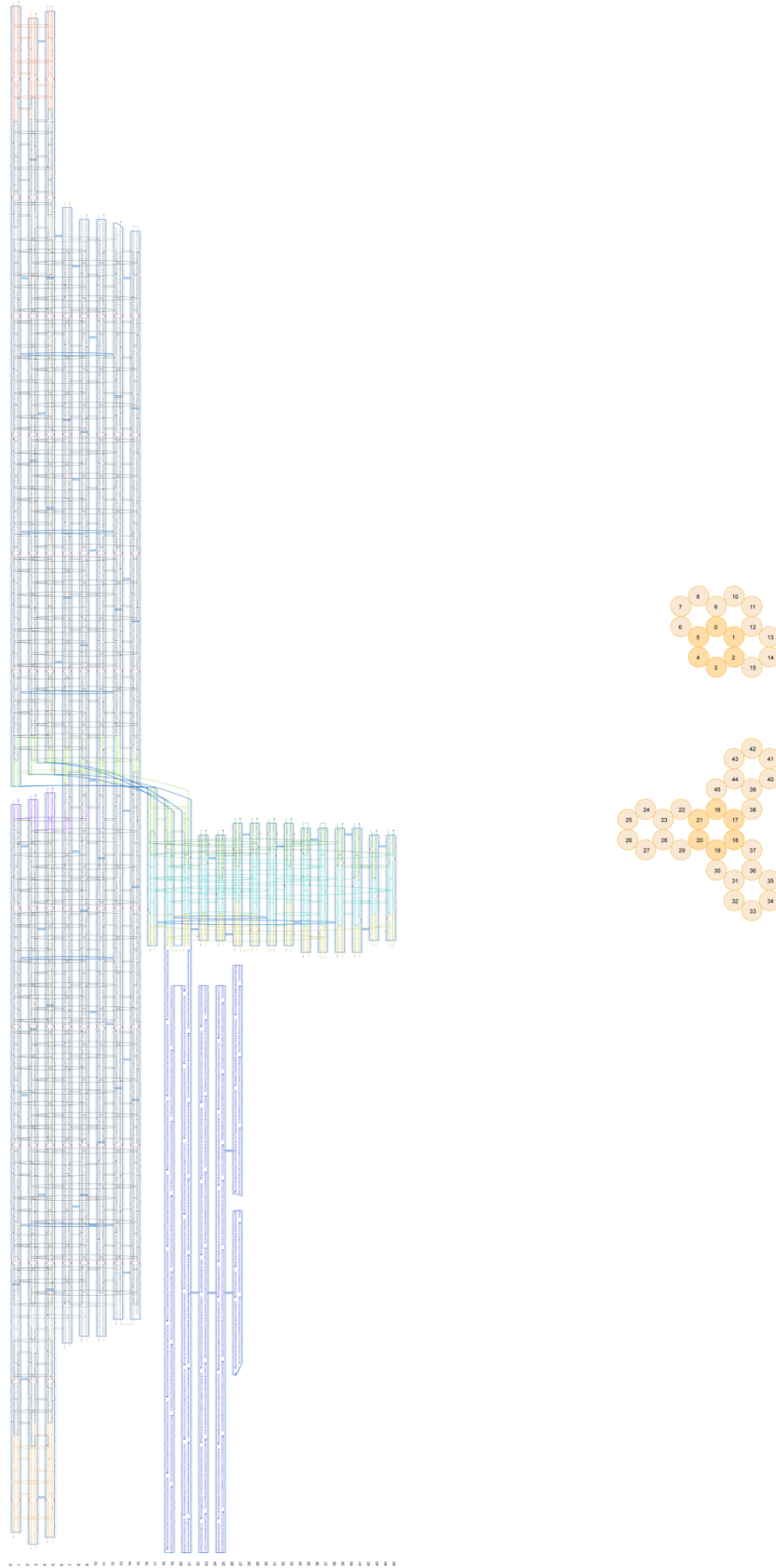

**Figure S5. Design diagram of the achiral turbine including load.**
This design was created with caDNAno *(28)*.

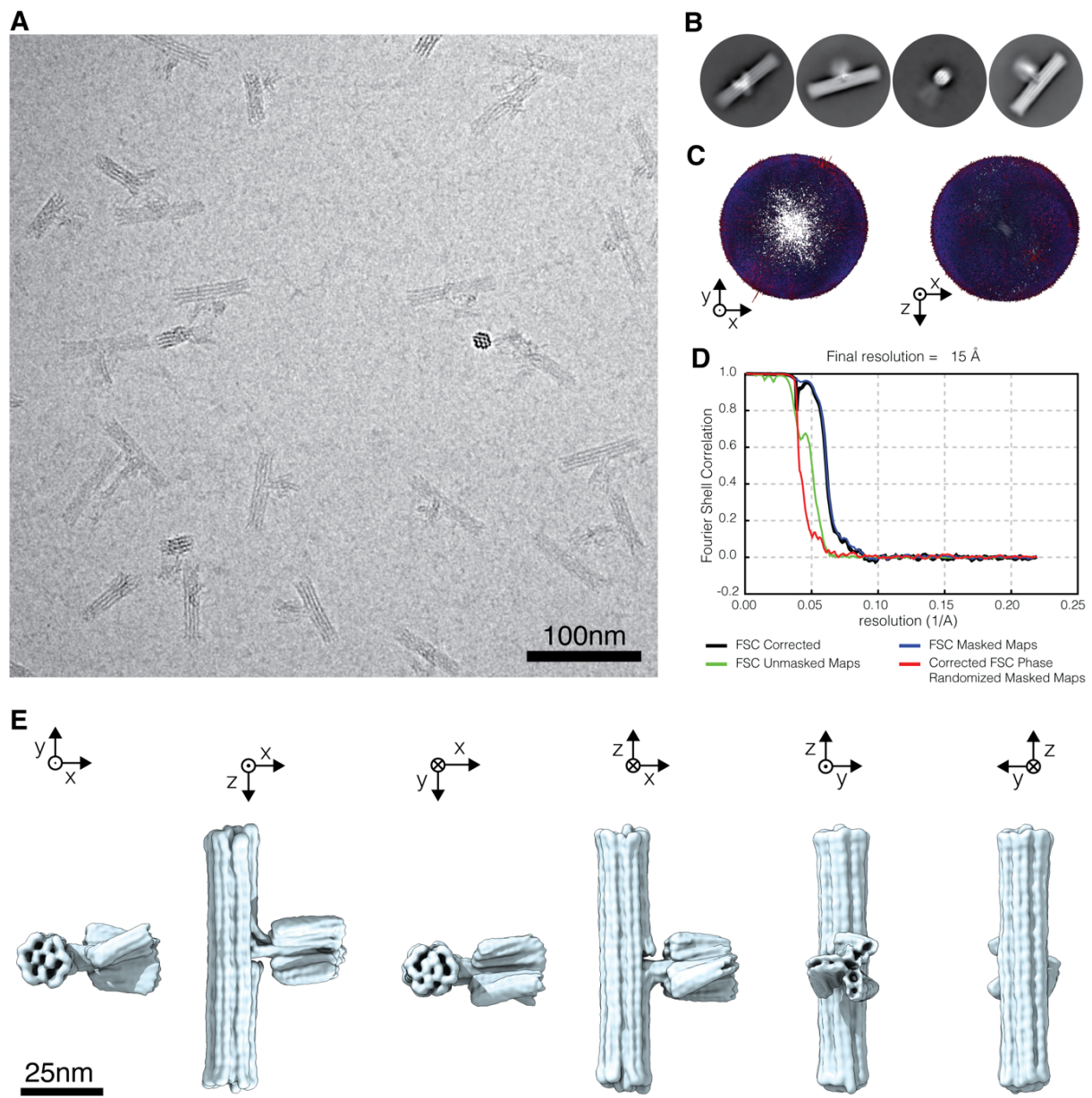

**Figure S6. Cryo-EM map determination for the left-handed turbine variant.**
**(A)** Exemplary micrograph of a total of 5,997 movie stacks. Scale bar is 100 nm. **(B)** Representative 2D class averages. **(C)** Histogram representing the orientational distribution of particles. **(D)** Fourier Shell Correlation (FSC) plot. **(E)** Six different views of the low-pass filtered electron density map. Scale bar is 25 nm.

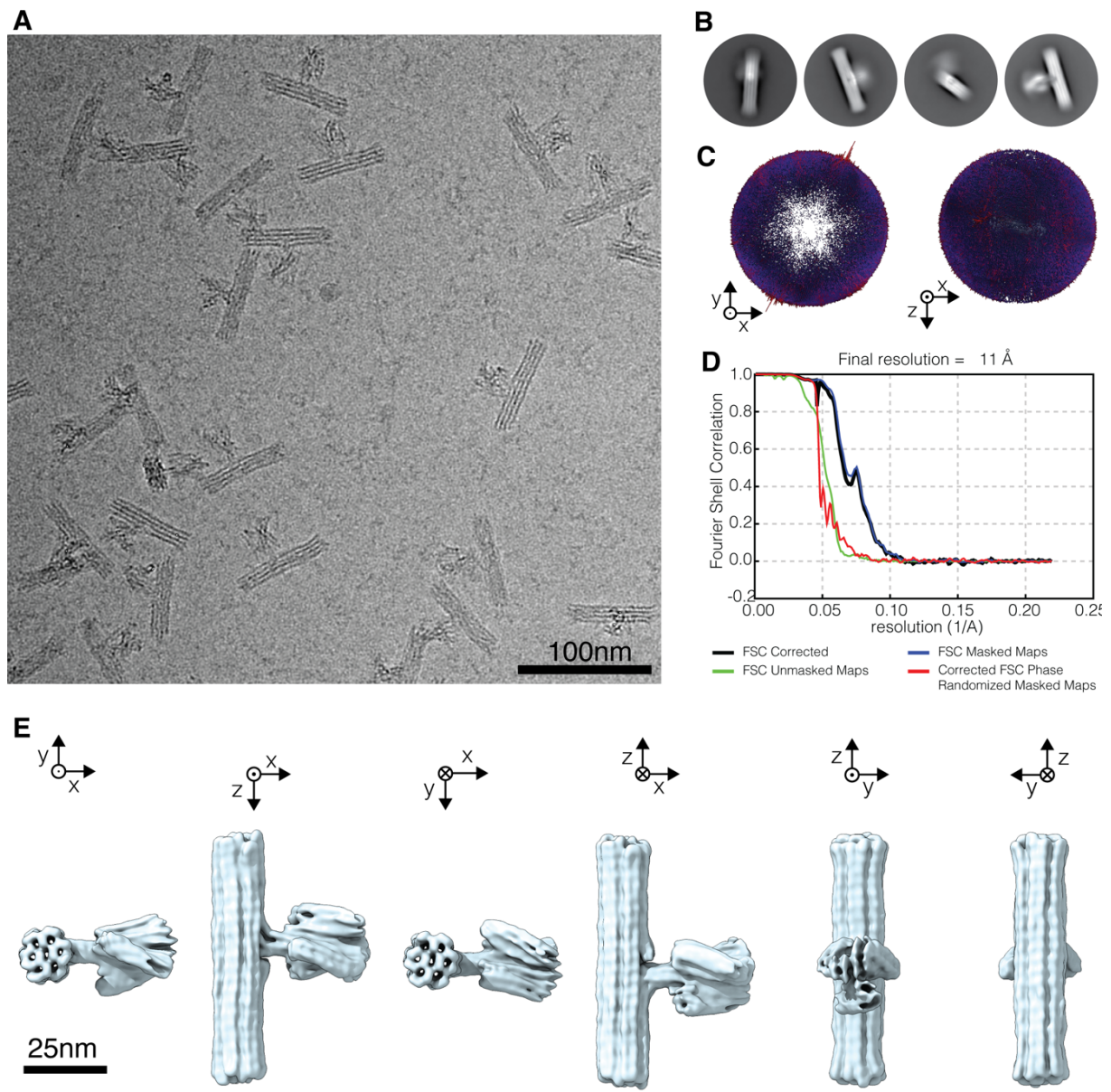

**Figure S7. Cryo-EM map determination for the right-handed turbine variant.**
**(A)** Exemplary micrograph of a total of 3427 movie stacks. Scale bar is 100 nm. **(B)** Representative 2D class averages. **(C)** Histogram representing the orientational distribution of particles. **(D)** Fourier Shell Correlation (FSC) plot. **(E)** Six different views of the low-pass filtered electron density map. Scale bar is 25 nm.

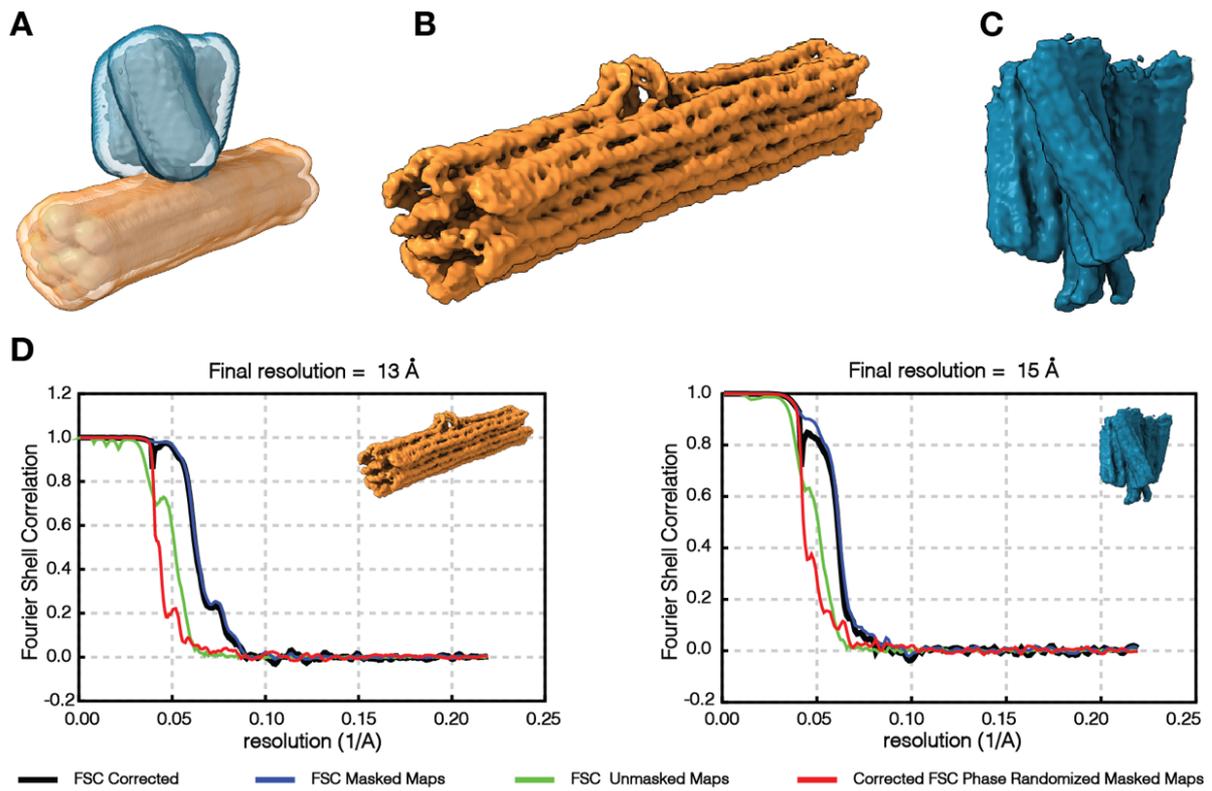

**Figure S8. Multibody (MB) refinement-based reconstruction of the left-handed turbine variant.**
**(A)** Masks of the rotor (blue transparent) and the lever (orange transparent) used for MB refinement. The map of the consensus refinement of the entire object is shown as a reference (gray). **(B)** Post-processed map of the MB-refined lever. **(C)** Post-processed map of the turbine reconstructed from a MB-refinement based set of partially-signal-subtracted particles. **(D)** FSC curves of the lever (left) and the turbine (right).

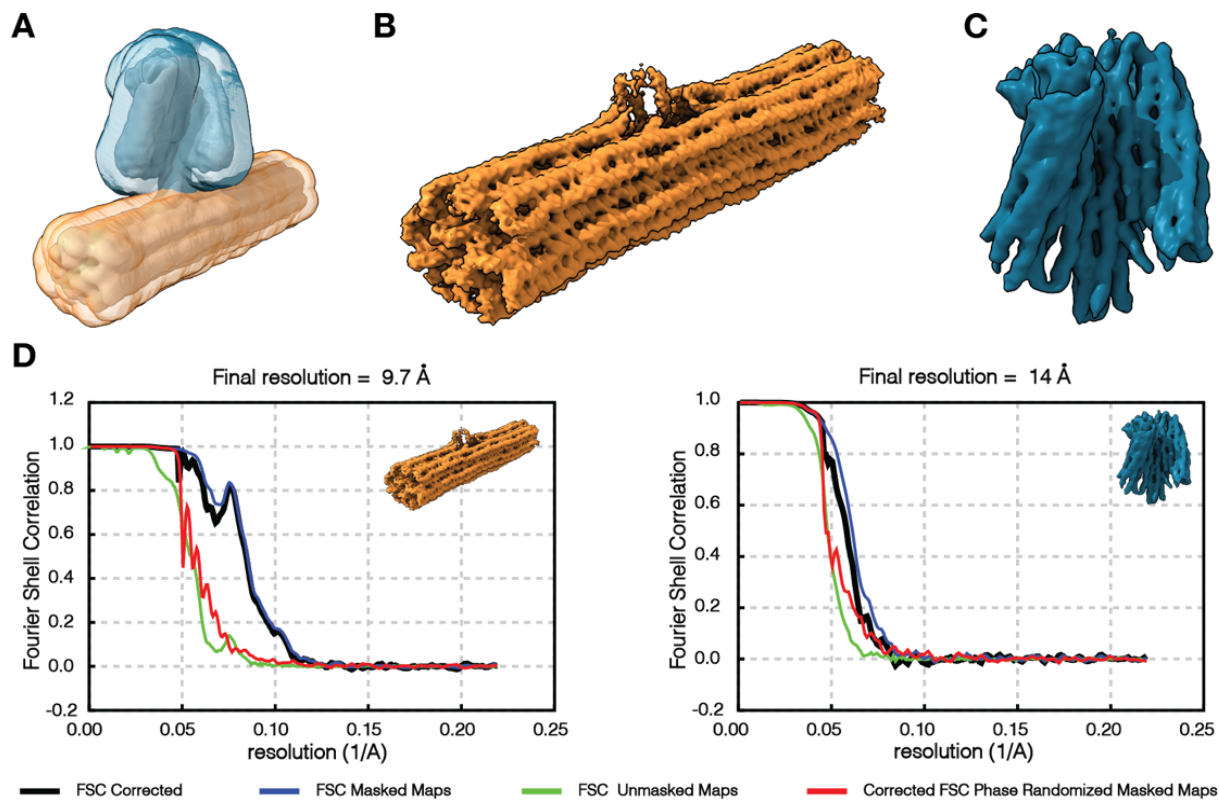

**Figure S9. Multibody (MB) refinement-based reconstruction of the right-handed turbine variant.**
**(A)** Masks of the rotor (blue transparent) and the lever (orange transparent) used for MB refinement. The map of the consensus refinement of the entire object is shown as a reference (gray). **(B)** Post-processed map of the MB-refined lever. **(C)** Post-processed map of the turbine reconstructed from a MB-refinement based set of partially-signal-subtracted particles. **(D)** FSC curves of the lever (left) and the turbine (right).

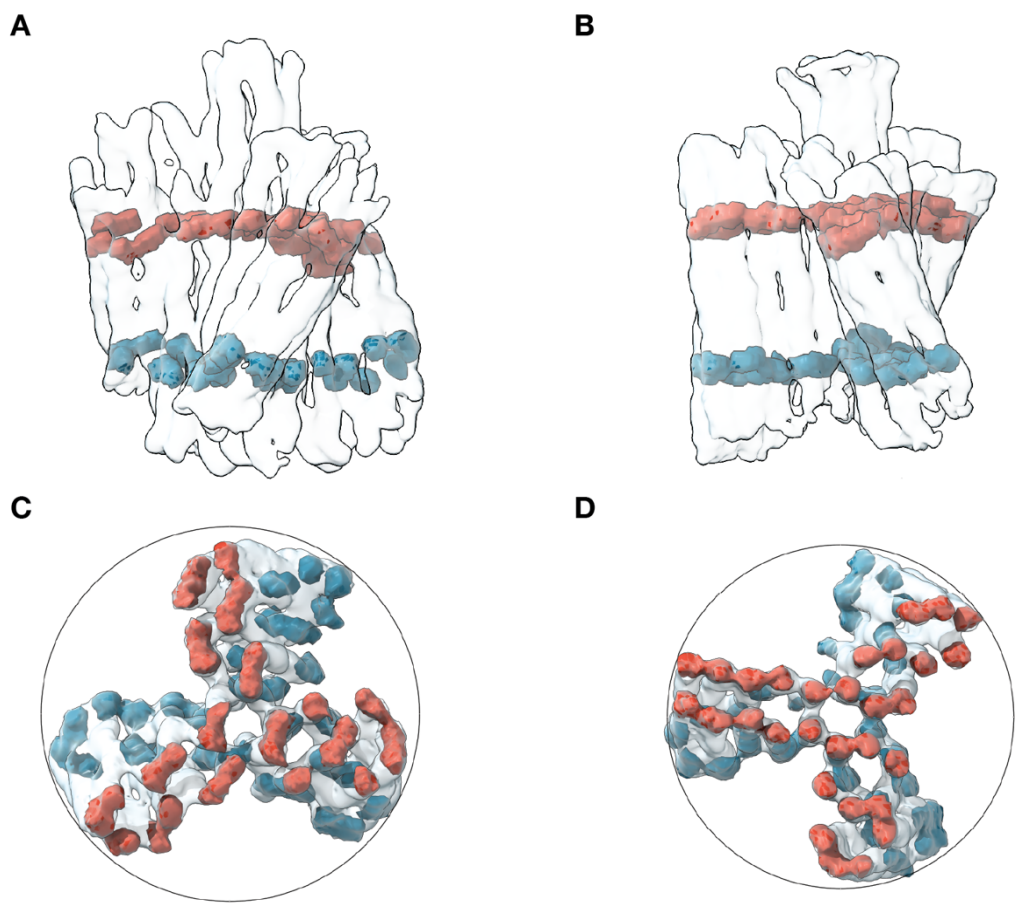

**Figure S10. Twist measurement.**
**(A)** Right-handed (RH) version of the turbine. The top (red) and bottom (blue) slice is extracted from the well-resolved central part of the map at base pair positions which are on the same plane in the design. **(B)** Like in A but for the left-handed (LH) version. **(C) – (D)** Top views of RH and LH versions. As the relative insertion density in the RH version is large than the relative deletion density in the LH version, the number of insertions is higher than the number of deletions (1 more vs 0.5 fewer bases compared to the regular 7 bp segments), and accordingly, a higher twist degree can be expected for the RH turbines.

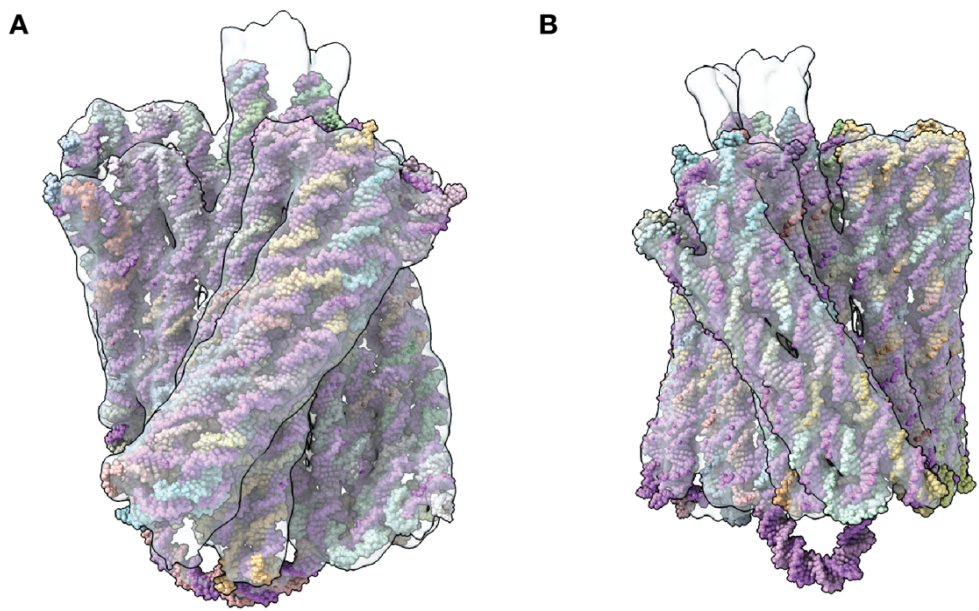

**Figure S11. Atomic model fits.**
**(A)** Right-handed version of the turbine. **(B)** Left-handed version of the turbine.

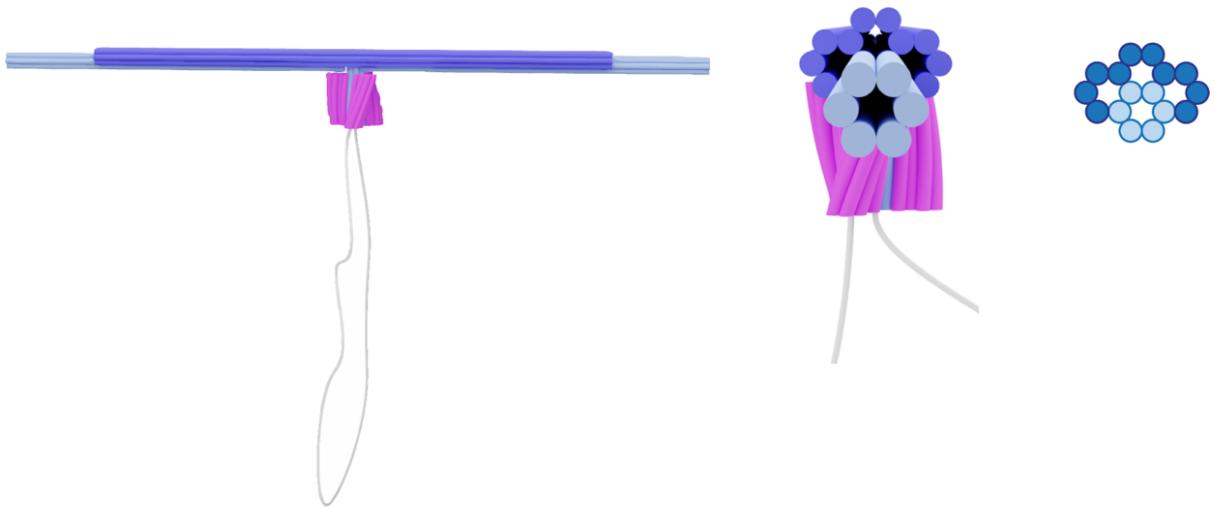

**Figure S12. Schematics of the front (left) and sideview (right) of the turbine with the DNA bundle.**
Note that middle 6hb part is reinforced with additional 10 helices (dark blue) to make it a 16hb structure.

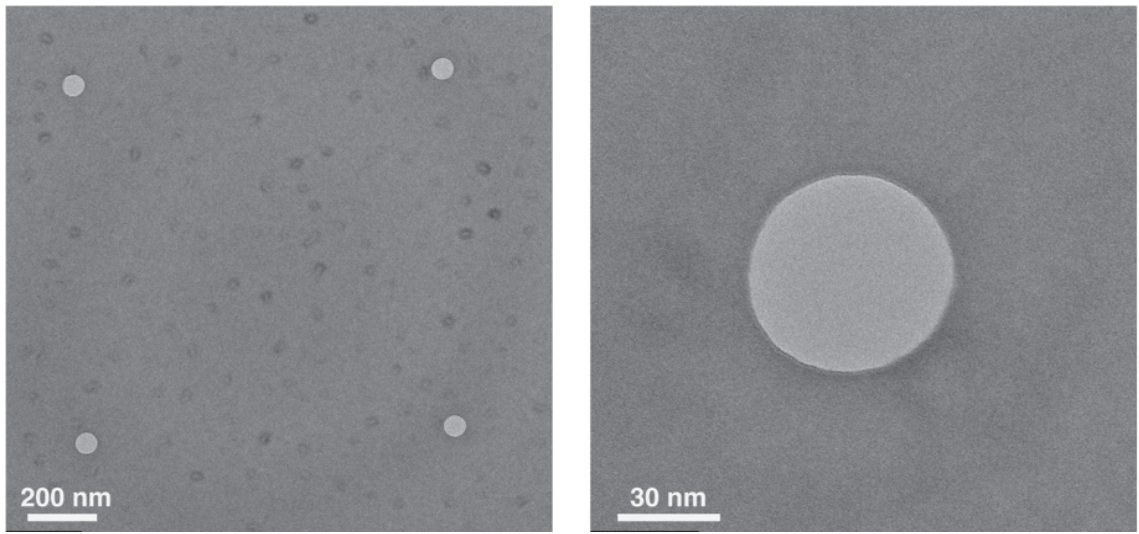

**Figure S13. Typical TEM images of solid-state nanopores used in the experiments.**

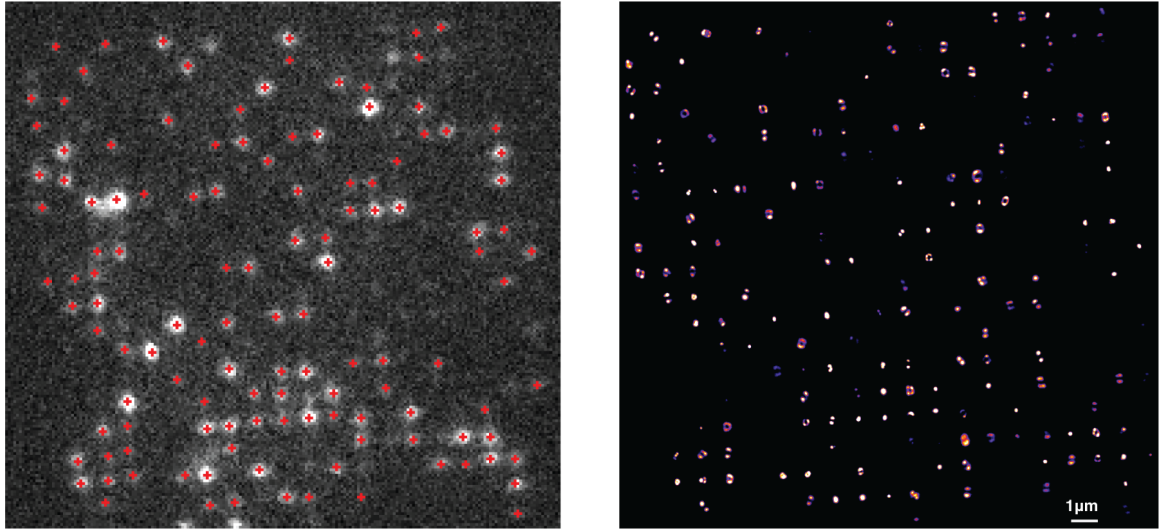

**Figure S14. Example of the high-throughput fluorescence imaging of rotors in a nanopore array.**
(Left) Example raw data frame of DNA turbines docked onto a nanopore array, and the corresponding single-particle localization results. Frame exposure time 5 ms. (Right) Results of the single-particle localization heatmap from 8000 stacked frames. Scalebar: 1 μm.

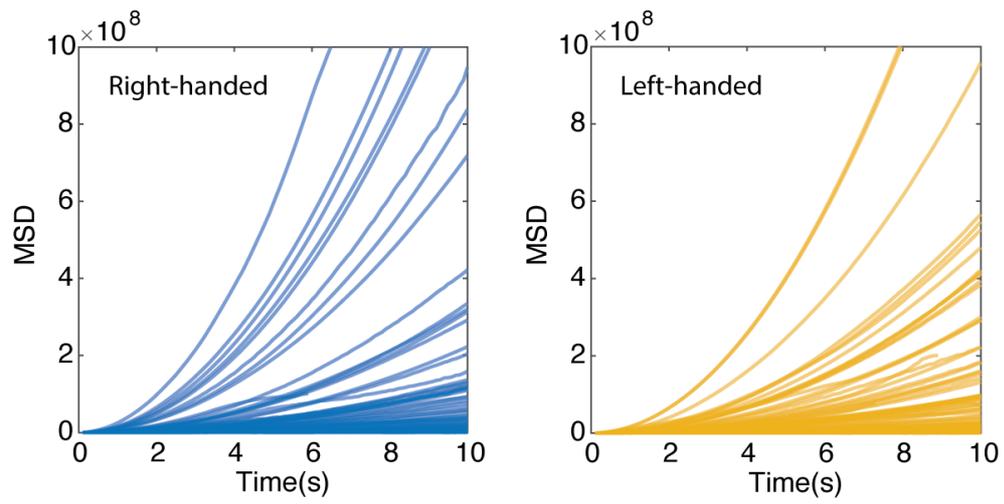

**Figure S15. MSD plots corresponding to the trajectories shown in Fig. 2.**

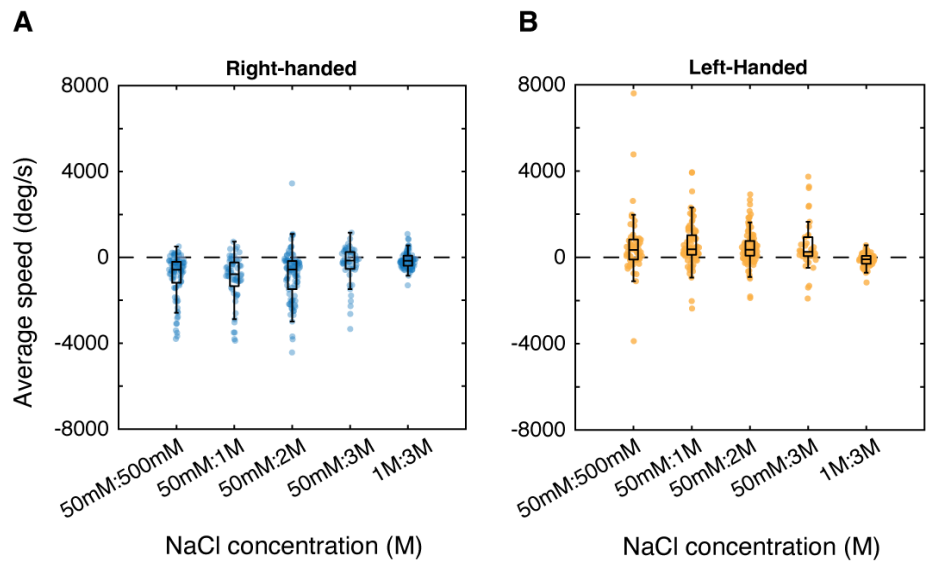

**Figure S16. Average rotation speed of DNA bundle driven left-handed and right-handed turbine.**
Data for different transmembrane NaCl concentration gradients, from left to right: 50mM:500 mM, 50mM:1M, 50mM:2M, 50mM:3 M, 1M:3 M ($n_A$= 124, 74, 116, 67, 154. $n_B$ = 98, 141, 159, 80, 130). In all box plots: center line, median; box limits, upper and lower quartiles; whiskers, 1.5x interquartile range.
.

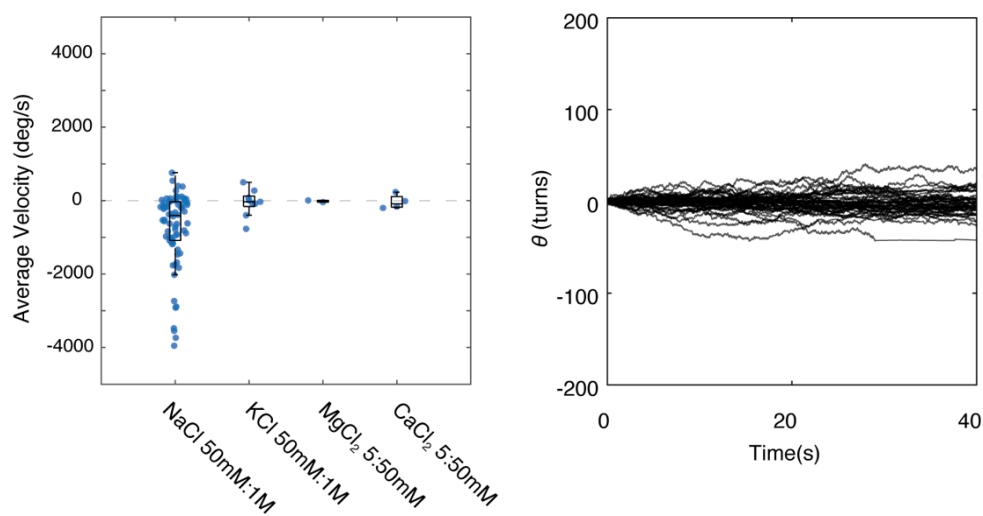

**Figure S17. Rotations under different types of cation gradients**.
Left: the calculated averaged velocity of turbines under 50mM:1M NaCl, 50mM:1M KCl, 5:50mM MgCl$_2$, and 5:50 mM CaCl$_2$, respectively (n = 74, 9, 2, and 4). Both divalent cation gradient conditions contain 50 mM KCl (equal concentration on both sides of the membrane). Right: example angular displacement of turbine motions under 5:50mM MgCl$_2$. In all box plots: center line, median; box limits, upper and lower quartiles; whiskers, 1.5x interquartile range.

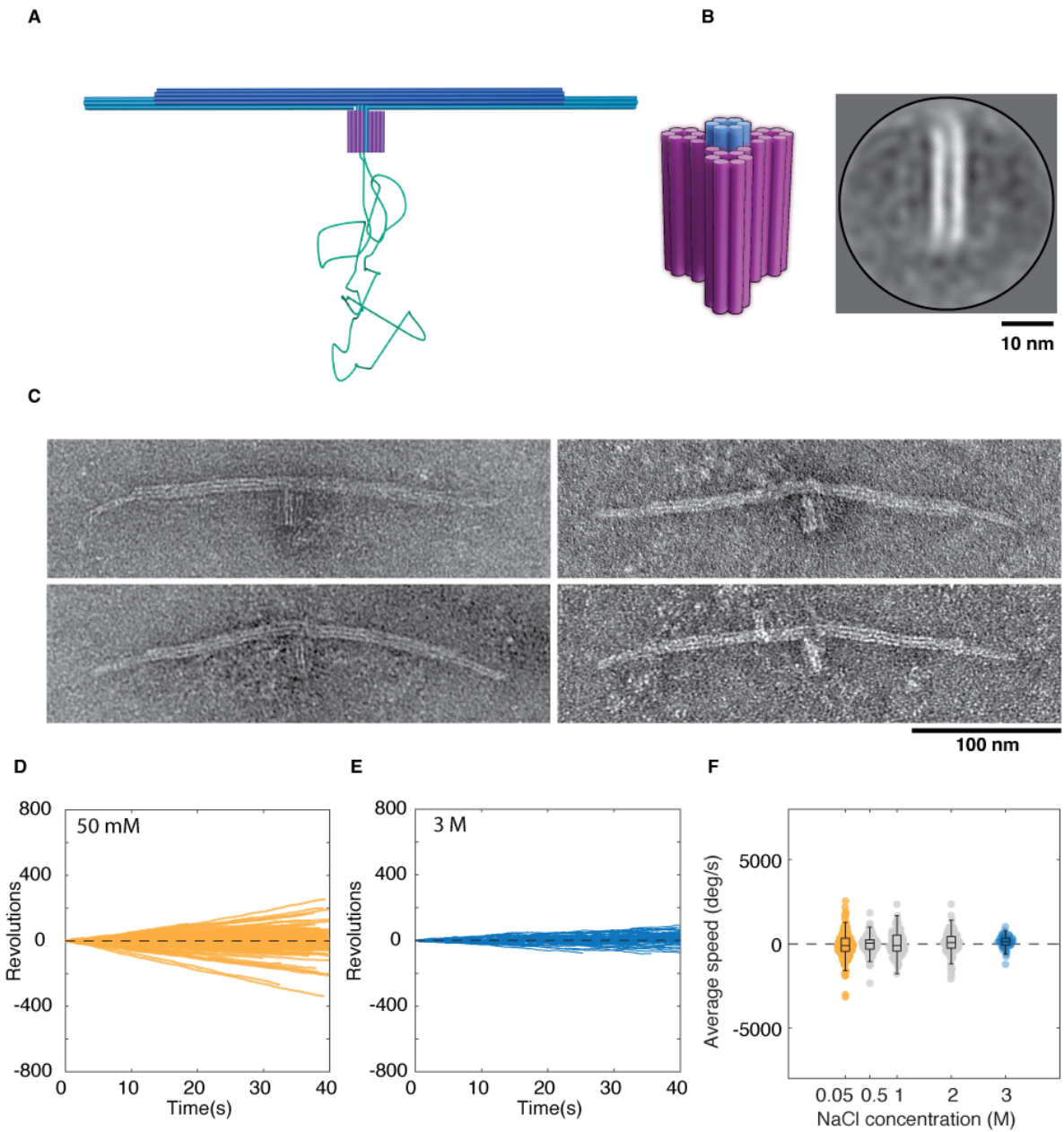

**Figure S18. Achiral turbines did not show a directional preference.**
**(A)** Schematic of the achiral turbine with load as a control structure. **(B)** Class-averaged negative stained TEM images of the achiral turbine structure with corresponding schematics. **(C)** Negative stained TEM images of the achiral turbine structure. **(D)** Example cumulative angular-displacement curves for achiral turbines under a 100-mV bias voltage in 50 mM NaCl (n = 259) and **(E)** in 3 M NaCl (n = 96). **(F)** Average rotation speed for achiral turbines in NaCl concentrations of 50 mM, 500 mM, 1M, 2M, 3M (n = 246, 250, 161, 173, and 90, respectively). In all box plots: center line, median; box limits, upper and lower quartiles; whiskers, 1.5x interquartile range.

**Left-handed**

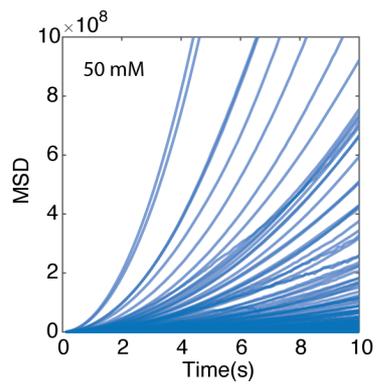 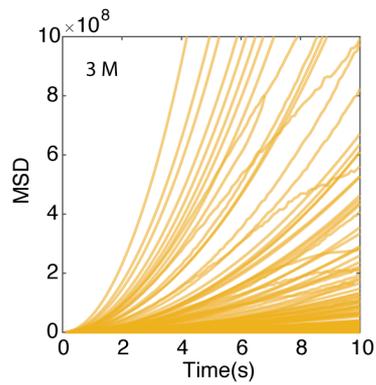

**Right-handed**

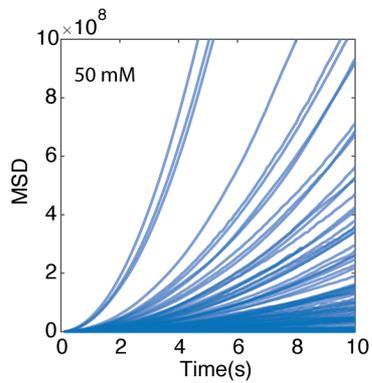 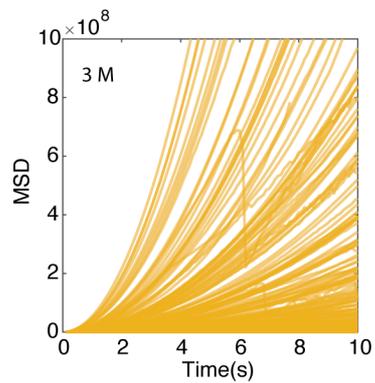

**Figure S19. MSD plots corresponding to the trajectories shown in Fig. 3.**

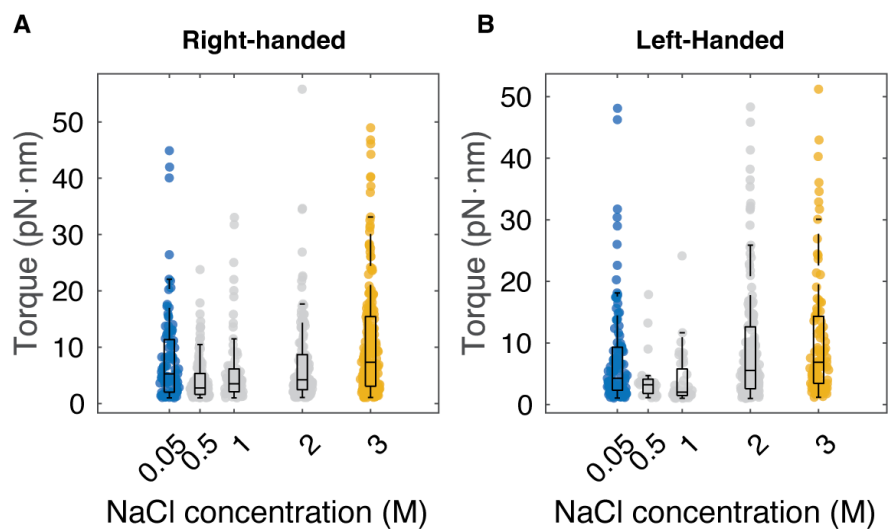

**Figure S20. Estimated torques for right-handed (A) and left-handed (B) DNA turbines.** All estimations result from the data shown in Fig. 3f and 3i (with the same sample sizes), obtained under 100 mV bias voltage. In all box plots: center line, median; box limits, upper and lower quartiles; whiskers, 1.5x interquartile range.

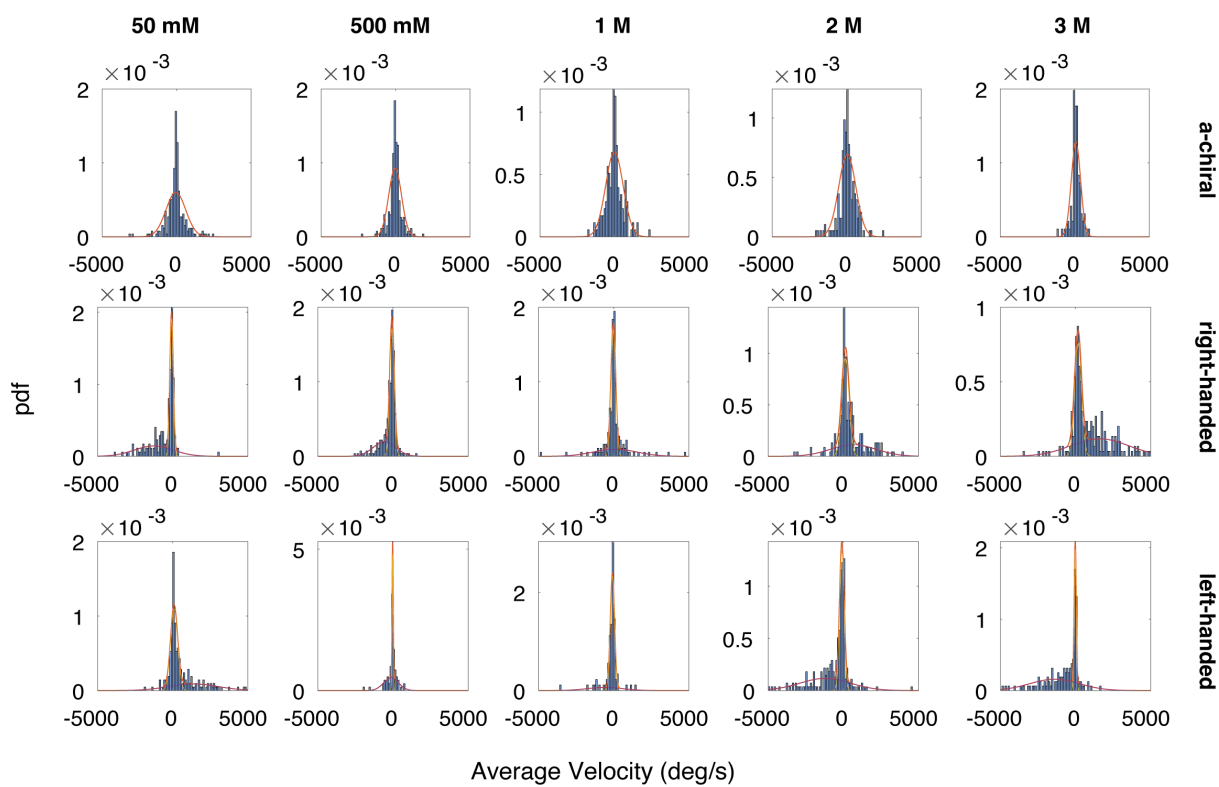

**Figure S21. Average velocity distribution and model fitting for rotation data from achiral (1st row), right-handed (2nd row), and left-handed (3rd row) structures.**
Each vertical column corresponds to one NaCl concentration (left to right): 50 mM, 500mM, 1M, 2M, 3M. For achiral structures (1st row), a single normal distribution was fitted to the data in all conditions. For chiral structures (2nd – 3rd rows), the sum of two normal distributions was fitted to the data, with one peak with a mean velocity of approximately zero (which are presumably due to nonfunctional structures because of nonspecific interaction with the surface, failure of the structural integrity, or incorrect docking), as well as one with a mean finite velocity.

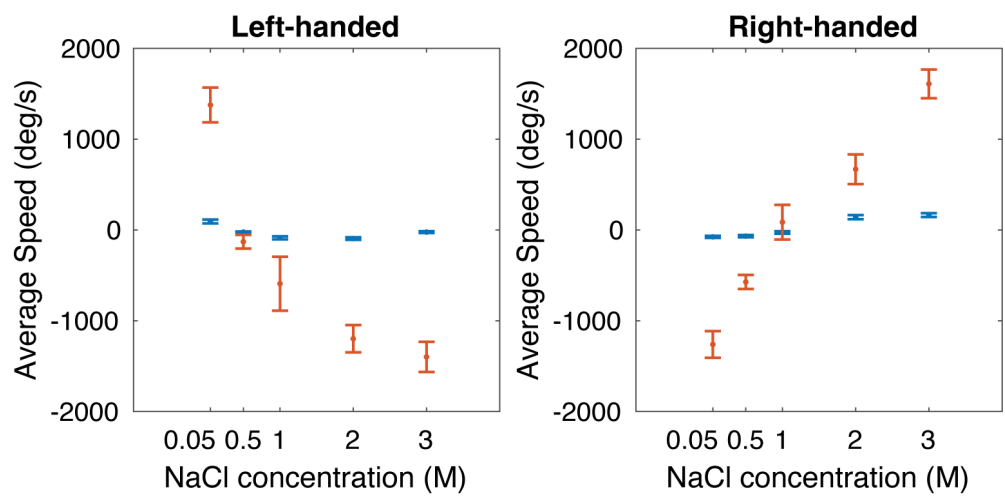

**Figure S22. Mean speeds of left- and right-handed structures from the fitting results in Fig. S20.**
The mean value of the peak near zero is marked in blue, and seen to be located near zero, as expected. The mean value of the second peak is marked in red (and shown in Fig. 3g and 3m in the main text, same sample sizes as Fig. 3f and 3l). Error bars indicate the standard error of the mean.

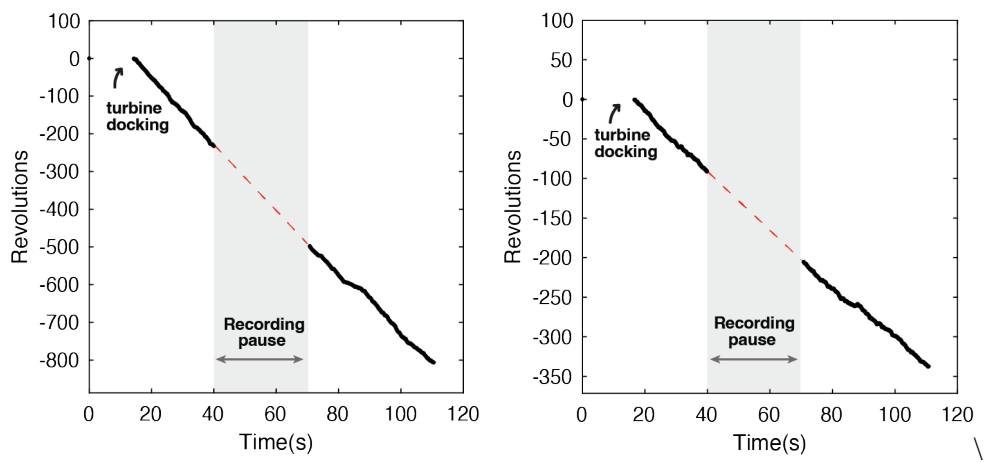

**Figure S23. Examples of rotations after 40 seconds.**
The recording was paused after the first 40-second period and restarted after 30 seconds. The starting Y value of each 2nd recording is the extrapolation of the first measurement segment (red dashed line). Data were recorded from the left-handed turbines, 3M NaCl, driven by 100 mV voltage.

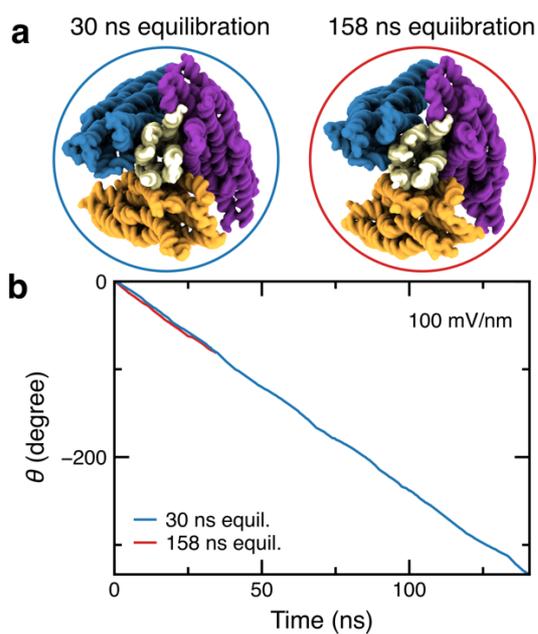

**Figure S24. Control simulation of alternative initial turbine configuration. (a)** For the results reported in the main text, the configuration of the turbine in 50 mM NaCl was frozen by an RMSD restraint (see Methods for details) after 30 ns of equilibration (left). A second system was constructed with RMSD restraint freezing the microscopic configuration of the rotor after 158 ns of equilibration (right). (**b**) Subject to an applied electric field, the two turbine configurations were found to rotate at a nearly identical rate.

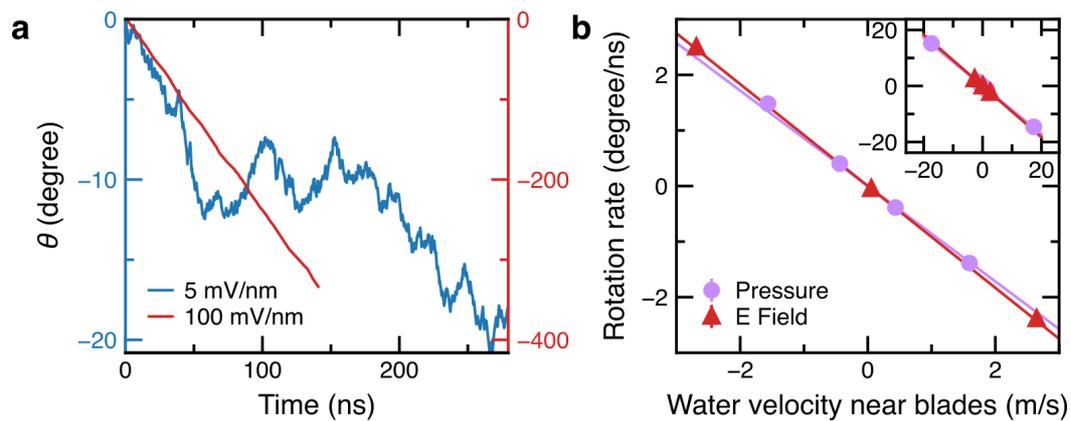

**Figure S25. Simulated rotation of turbine in 50 mM NaCl under various conditions.**
**(a)** Rotation of the turbine at 5 mV/nm (blue; left axis) or 100 mV/nm (red; right axis) applied bias. **(b)** Rotation rate versus local water velocity near the turbine.

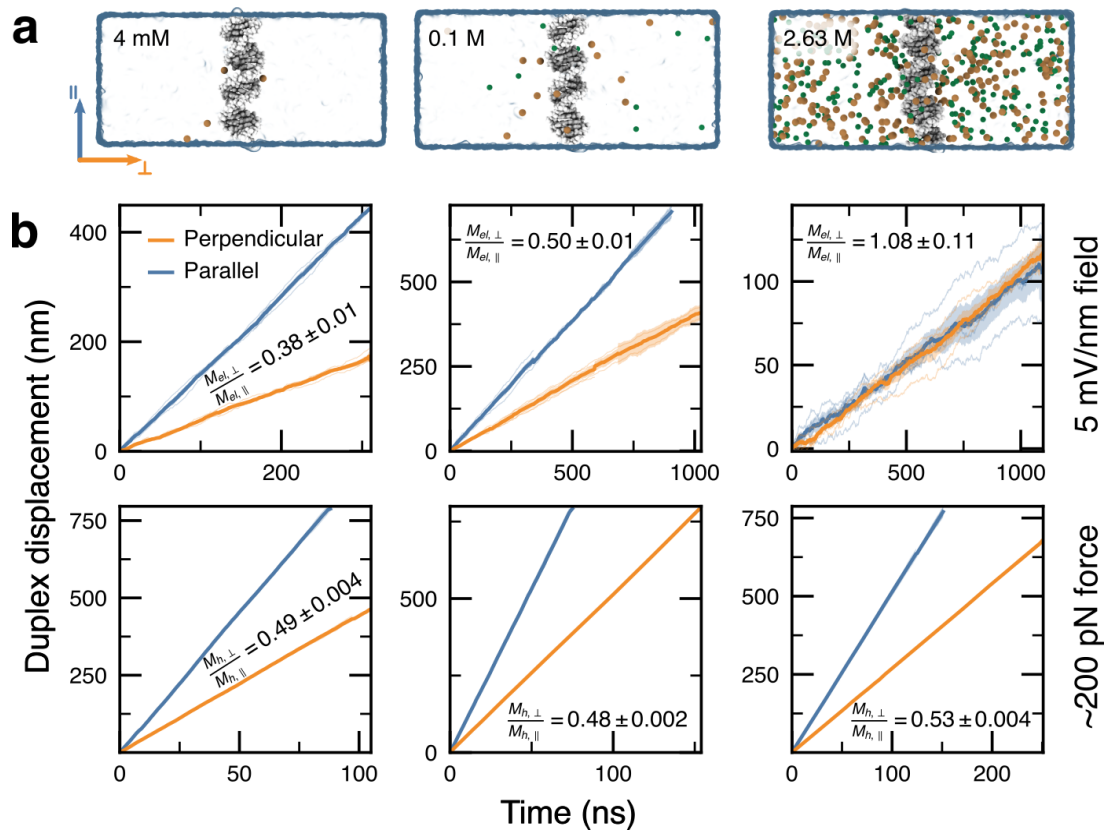

**Figure S26. Dependence of DNA helix mobility on salt concentration in MD simulations.**
(**a**) Three systems containing a single DNA helix in electrolyte solution with 10% of ions depicted. (**b**) Center-of-mass displacement of the DNA helix relative to the water when driven by an electric field (top row) or force (bottom row) applied parallel (blue) or perpendicular (orange) to the DNA axis. For each condition, four simulations were performed (thin lines), allowing the average displacement (thick lines) and the standard error of the mean (shaded region) to be computed at each frame of the trajectory. The mobility was estimated for each replicate simulation from a linear regression, yielding the annotated perpendicular-to-parallel mobility ratio and the standard error. Top row: a 5 mV/nm field drove displacement. Bottom row: a force applied to the DNA (104, 105, and 114 pN in systems representing 0.004 , 0.1, and 2.6 M conditions, respectively) with an equal and opposite force distributed over the oxygen atoms of the water. A momentum-conserving Lowe-Andersen thermostat held the temperature near 295 K during the simulations.

**Supplementary Movie S1:** MD simulation of the right-handed DNA turbine rotation in 100 mM NaCl condition under electric field, as shown in Fig 4.

**Supplementary Movie S2:** MD simulation of the right-handed DNA turbine rotation in 3 M NaCl condition under electric field, as shown in Fig 4.

**Supplementary References**